\documentclass{amsart}

\usepackage{amsmath,amssymb,amsthm,amsfonts}
\usepackage{latexsym}
\usepackage{mathrsfs}
\usepackage{bbm}

\usepackage{tabularx}
\newcommand{\PreserveBackslash}[1]{\let\temp=\\#1\let\\=\tem}

\numberwithin{equation}{section}

\newcommand{\p}{\partial}
\newcommand{\dbar}{\bar\partial}
\renewcommand{\d}{\mathrm{d}}

\newcommand{\IC}{\mathbbm{C}}
\newcommand{\IM}{\mathbbm{M}}
\newcommand{\IP}{\mathbbm{P}}
\newcommand{\IR}{\mathbbm{R}}
\newcommand{\IT}{\mathbbm{T}}

\newcommand{\CP}{\mathbbm{C}\mathbbm{P}}

\newcommand{\cD}{\mathscr{D}}
\newcommand{\cF}{\mathscr{F}}
\newcommand{\cH}{\mathscr{H}}
\newcommand{\cL}{\mathscr{L}}
\newcommand{\cM}{\mathscr{M}}
\newcommand{\cN}{\mathscr{N}}
\newcommand{\cO}{\mathscr{O}}
\newcommand{\cP}{\mathscr{P}}
\newcommand{\cS}{\mathscr{S}}
\newcommand{\cT}{\mathscr{T}}
\newcommand{\cU}{\mathscr{U}}

\newcommand{\CL}{\mathcal{L}}
\newcommand{\CN}{\mathcal{N}}

\newcommand{\hook}{{\setlength{\unitlength}{12pt}%adjust point size
     \begin{picture}(.833,.8)
       \put(.15,.08){\line(1,0){.35}}
       \put(.5,.08){\line(0,1){.5}}
       \end{picture}}}

\newtheorem{thm}{Theorem}

\newtheorem{defn}{Definition}
\newtheorem{lemma}{Lemma}

\begin{document}

\title[Twistor Actions for Self-Dual Supergravities]{
Twistor Actions for Self-Dual Supergravities}

\begin{flushright}
  {\small Imperial--TP--MW--02/07}
  \vspace*{1cm}
\end{flushright}

\author{Lionel J. Mason}

\address{The Mathematical Institute, University of Oxford
24--29 St.~Giles, Oxford OX1 3LP, United Kingdom}
\email{lmason@maths.ox.ac.uk}

\author{Martin Wolf}
\address{Theoretical Physics Group, The Blackett Laboratory,
Imperial College London,
Prince Consort Road, London SW7 2AZ, United Kingdom}
\email{m.wolf@imperial.ac.uk}

\date{June 13, 2007}

\begin{abstract}
  We give holomorphic Chern-Simons-like action functionals on
  supertwistor space for self-dual supergravity theories in four dimensions,
  dealing with $\CN=0,\ldots,8$ supersymmetries, the cases where
  different parts of the $R$-symmetry are gauged, and with or without
  a cosmological constant.  The gauge group is formally the group of
  holomorphic Poisson transformations of supertwistor space where the
  form of the Poisson structure determines the amount of $R$-symmetry
  gauged and the value of the cosmological constant.  We give a
  formulation in terms of a finite deformation of an integrable
  $\dbar$-operator on a supertwistor space, i.e., on regions in
  $\CP^{3|8}$.  For $\CN=0$, we also give a formulation that does not
  require the choice of a background.
\end{abstract}

\thanks{}

\maketitle
\tableofcontents

\section{Introduction}

Recently it has been discovered that $\CN=8$ supergravity has better
ultraviolet behaviour than has hitherto been anticipated,
Bjerrum-Bohr, et\ al. (2007), Bern et\ al.\ (2007) and Green et\ al.\
(2007).  This has  led some authors to speculate that it is possibly even
finite.  This improved behaviour relies on exact cancellations that do
not follow from standard supersymmetry arguments, Stelle (2005).  One
possible explanation arises from twistor string theory, Witten (2003)
and Berkovits (2004).  The original twistor string theories by Witten
and Berkovits correspond to conformal supergravity (together with
supersymmetric Yang-Mills theory), Berkovits \& Witten (2004).  By
gauging certain symmetries of the Berkovits twistor string, Abou-Zeid,
Hull \& Mason (2008) introduced a new family of twistor string
theories some of which have the appropriate field content for Einstein
supergravity (including $\CN=4$ and $\CN=8$).  Such a twistor string
formulation of Einstein supergravity could be an explanation for the
possible ultraviolet finiteness of $\CN=8$ supergravity if it were
fully consistent in its quantum theory.  However, it now appears that
these twistor string theories are chiral, Nair (2008), unlike the
the original twistor string theories which were parity invariant.  It
remains a major open question as to whether a twistor-string theory exists
that gives the full content of Einstein (super)-gravity even just at
tree level.  

An approach to understanding what the appropriate twistor string
theory might be is via a twistor action, Mason (2005) and Boels, Mason
\& Skinner (2007a,b).  Such actions have two terms.  The first on its
own gives a kinetic term for all the fields, but with only the
self-dual part of the interactions.  The second gives the remaining
interactions of the full theory and correspond to the instanton
contribution in the twistor-string theory.  In the case of $\CN=4$
supersymmetric Yang-Mills theory, the self-dual part of the action on
twistor space is a holomorphic Chern-Simons theory, Witten (2004), see
also Sokatchev (1995) for a closely related harmonic superspace
action.  Berkovits \& Witten (2004) gave a twistor action for
self-dual $\CN=4$ conformal supergravity.  The purpose of this paper is to
give an analogous action in the case of self-dual $\CN=8$ Einstein
supergravity.  This action is special to $\CN=8$ supergravity in much
the same way as Witten's Chern-Simons action is special to $\CN=4$
supersymmetric Yang-Mills theory.  It lends general support to the
idea that twistor space has something special to say about full
$\CN=8$ supergravity and is suggestive of the existence of an
underlying twistor string theory, perhaps even with explicit $\CN=8$
supersymmetry as opposed to those of Abou-Zeid et\ al.\ (2008) in
which only $\CN=4$ supersymmetry is manifest.

Penrose's non-linear graviton construction (1976) reformulates the
local data of a four-metric with self-dual Weyl tensor into the
complex structure of a deformed twistor space, a three-dimensional
complex manifold obtained by deforming a region in $\CP^3$.  The
space-time field equation in this case is the vanishing of the
anti-self-dual part of the Weyl tensor, and in the Atiyah, Hitchin \&
Singer (1978) approach to twistor theory, this is reformulated as the
integrability of the twistor almost complex structure.  Berkovits \&
Witten (2004) introduce a
version of conformal gravity with just self-dual interactions in which
the underlying conformal structure is self-dual, but in which there is
also a linear
anti-self dual conformal gravity field (a linearised anti-self-dual
Weyl tensor $B$) propagating on the self dual background.  This has a
Lagrange multiplier 
action  (analagous to a `BF' action) 
$$
\int (B ,C^-)  \, {\mathrm {d\, vol}}\, ,
$$ 
where $C^-$ is the anti-self-dual part of the Weyl tensor, and
$(B,C^-)$ is the natural pairing.  This can be extended to $\CN=4$
supersymmetry.  Berkovits \& Witten (2004) gave a corresponding
(supersymmetric) twistor action of the form $\int b N$ where $N$ is
the Nijenhuis tensor of the almost complex structure and $b$ is a
Lagrange multiplier that doubles up as the Penrose transform of the
field $B$ when the field equations are satisfied.  In the
non-supersymmetric case, this was extended to a twistor action for
full (non-self-dual) conformal gravity in Mason (2005) with further
supersymmetric extension and connections with twistor-string theory
in Mason \& Skinner (2008).

For Einstein gravity we wish to encode the vanishing of the Ricci
tensor.  In the non-linear graviton this can be characterised by
requiring that the twistor space admits a fibration over a $\CP^1$
together with a certain Poisson structure up the fibre.  Ward (1980)
extended this to the Einstein case, with a cosmological constant; in
this case, the twistor space is required to admit a holomorphic
contact structure that is non-degenerate when the cosmological
constant is non-zero, see Ward \& Wells (1990) and Mason \& Woodhouse
(1996) for textbook treatments.  So, for Einstein gravity, we are
seeking a twistor action whose field equations not only imply the
integrability of an almost complex structure, but also the existence
of some compatible holomorphic geometric structure, for example the
contact one-form in the case of the cosmological constant, or the
fibration together with a Poisson structure up the fibres in the case
of vanishing cosmological constant.  The first task is to introduce
suitable variables that encode the almost complex structure together
with the relevant compatible geometric structure on the real
six-manifold underlying the twistor space.  This turns out to be a 
one-form with values in a line bundle, and we write down the appropriate field
equations that it must satisfy and an action (depending also on a
Lagrange multplier field) that gives rise to them; the Lagrange
multiplier field again corresponds to an anti-self-dual linear
gravitational field propagating on the self-dual background via the
Penrose transform when the field equations are satisfied.

Our primary exposition will focus on the $\CN=8$ supersymmetric cases,
and reduce them to the cases with lesser or no supersymmetry.
Supersymmetric extensions of Penrose's non-linear graviton
construction were first discussed by Merkulov (1992a,b) (see also
Merkulov (1991,1992c)) based on work by Manin (1988) and developed
further in Abou-Zeid, Hull \& Mason (2008) and in Wolf
(2007).\footnote{See also S\"amann (2006) and Wolf (2006) and
  references therein for recent reviews of supertwistors and their
  application to supersymmetric gauge theories.}  That in Wolf (2007)
gives a twistor description of four-dimensional $\CN$-extended,
possibly gauged, self-dual supergravity with and without cosmological
constant in terms of a deformed supertwistor space, a deformation of a
region in $\CP^{3|\CN}$ endowed with an even holomorphic contact
structure.  Here we also discuss the different gaugings in the case
without a cosmological constant.  It is these the integrability of the
almost complex structures of these twistor spaces together with the
holomorphy of the appropriate geometric structures that correspond to
the field equations for our twistor actions.

There are now a number of contexts arising from conventional string
theory and M-theory in which the task of finding
variables and action principles whose field equations encode the
integrability of complex structures compatibly with some other
geometric structure.  In particular Kodaira-Spencer theory, Bershadsky
et. al. (1994), leads to field equations that imply the integrability
of an almost complex structure compatible with a global holomorphic
volume form on a six-manifold, yielding a Calabi-Yau structure.  For a
compendium of such theories and relations between them, including
conjectured relations to twistor-string theory see Dijkgraaf et.\ al.\
(2004).  The situations considered here are distinct from those in
Dijkgraaf et.\ al.\ (2004), but given that one of the form theories
involved there is a self-dual form theory of four-dimensional gravity
including a cosmological constant (see also Abou-Zeid \&
Hull (2006)) there may well be some important connections between these
ideas.

The paper is structured as follows. In \S\ref{twistorcon}, we first
review the equations of self-dual supergravity, with cosmological
constant and gauged $R$-symmetry, and then go on to review the various
twistor constructions and give a brief proof of the version of the
non-linear graviton construction for self-dual Einstein supergravity
both with and without cosmological constant and different gaugings.
In \S\ref{twistoraction}, we study infinitesimal deformations and show
that a deformation of the contact structure determines a deformation
of the almost complex structure.  We develop a non-projective twistor
formulation that shows that this persists in the case of a finite
deformation giving a compact form for the field equations, i.e., the
integrability condition for the almost complex structure.  In the case
of maximal supersymmetry, $\CN=8$, we present the twistor action and
show that it gives the appropriate field equations.  We give a brief
discussion of its invariance properties and various reductions with
lesser gauging, supersymmetry, or no cosmological constant.

A Chern-Simons action is always expressed in a given background frame
and is not manifestly gauge invariant.  In this gravitational context,
our action is similarly not manifestly diffeomorphism invariant; we
require the choice of some background, which we take to be a solution
to the field equations.  However, we go some way towards an invariant
formulation.  We give an invariant formulation of the field equations
in general, but only find an explicitly diffeomorphism invariant
action in the $\CN=0$ case with cosmological constant.
We
prove that on any smooth manifold of dimension $4n+2$ equipped with a
complex one-form $\tau$ up to scale (i.e., a complex line subbundle of
the complexified cotangent bundle), then, if $\tau\wedge
(\d\tau)^n=0$, and a non-degeneracy condition is satisfied, there is a
unique integrable almost complex structure for which $\tau$ is
proportional to a non-degenerate holomorphic contact structure. This
idea can be used to give a covariant form of the field equations in
general, and a covarant action in the $\CN=0$ case.

In \S\ref{conclusions}, we make some general concluding remarks.  An
action principle for $\CN=8$ self-dual supergravity with vanishing
cosmological constant has been obtained by Karnas \& Ketov (1998) in
harmonic superspace for split space-time signature.\footnote{A similar
  action for $\CN=4$ supersymmetric Yang-Mills theory was discovered
  in the context of harmonic superspace by Sokatchev (1995).}  In that
work, harmonic superspace is the spin bundle of super space-time and
in Euclidean signature, it can naturally be identified with the
supertwistor space.  However, their action uses structures pulled back
from space-time (e.g., the Laplacian) that are not locally obtainable
from the complex structure and contact structure on twistor space.  It
is therefore not possible to regard it as a twistor action.
Nevertheless, their action is closely related to ours and we show that
theirs can be obtained from ours by gauge fixing in appendix
\ref{sec:app-pf}.  In appendix \ref{nonproj} we give a detailed
discussion of the construction of the line bundle on a super-twistor
space whose total space corresponds to a non-projective twistor space.
In the appendix \ref{sec:app-superbf}, we discuss some alternative
twistor actions.

\section{Twistor constructions for self-dual
supergravity}\label{twistorcon}

We work throughout in a complex setting.  This can be understood as
arising from taking a real analytic metric on a real space-time, and
extending it to become a holomorphic complex metric on some
neighbourhood $M$ of the real slice in complexified space-time.  We
can straightforwardly restrict attention to Euclidean or split
signature slice by requiring invariance under appropriate
anti-holomorphic involutions (for Euclidean signature,
these are discussed in the appendix \ref{sec:app-pf}).
In the Euclidean case, one needs to restrict the number of
allowed supersymmetries $\CN$ to be even.

\subsection{Definitions, notation and conventions}
We model our definition of chiral super space-time 
on the paraconformal geometries of Bailey \& Eastwood (1991) (see
also Wolf (2007))

\begin{defn}
  A {\it right-chiral} super space-time, $\cM$, is a split
  supermanifold of super-dimension $4|2\CN$ on which we have an
  identification\footnote{By $T\cM$ we will mean $T^{(1,0)}\cM$.
    There will be no role for anti-holomorphic objects on $\cM$.}
  $T\cM\cong\cH\otimes\widetilde\cS$, where $\widetilde\cS$ is the
  right (dotted) spin bundle of rank $2|0$ and $\cH$ is the sum of the
  left spin bundle $\cS$ and the rank-$0|\CN$ bundle of supersymmetry
  generators and so has rank $2|\CN$.  We will also assume that $\cS$
  and $\cH$ are endowed with choices of Berezinian forms (so that
  $T\cM$ does also).
\end{defn}

This is the superspace one
would obtain from a full super space-time by eliminating the
left-handed fermionic coordinates, leaving only the right-handed ones in play.
Being a split supermanifold, it is locally of the form
$\mathbbm{C}^{4|2\CN}$ with coordinates\footnote{The index structure
   on the bosonic coordinates in the curved case is not natural, but
   simplifies notation.}  $(x^{\mu\dot\nu},
\theta^{m\dot\nu}):=x^{M\dot\nu}$ with $x^{\mu\dot\nu}$ bosonic and
$\theta^{m\dot\nu}$ fermionic where the indices range as follows:
$\alpha,\ldots,\mu,\ldots=0,1$ for left-handed two-component spinors,
$\dot\alpha,\ldots,\dot\mu,\ldots=\dot 0,\dot1$ for right-handed
spinors, $i,\ldots,m,\ldots=1,\ldots,\CN$ indexing the supersymmetries
and $A=(\alpha,i)$, $M=(\mu,m)$; it will turn out in the following that it
is natural, and simplifying in this self-dual context to group
together the supersymmetry index $m$ and the undotted spinor index
$\mu$ into one index $M$.  We use the convention that letters
from the middle of the alphabets are coordinate indices whereas
letters from the beginning of the alphabets are structure frame
indices.  

The identification $T\cM\cong\cH\otimes\widetilde\cS$ will be specified
by a choice of `structure co-frame' given by the indexed one-forms
\begin{equation}
 E^{A\dot\alpha}\ =\ \d x^{M\dot\nu}{E_{M\dot\nu}}^{A\dot\alpha}.
\end{equation}
The dual vector fields will be denoted $E_{A\dot \alpha}$, $E_{A\dot
  \alpha}\hook  E^{B\dot\beta}={\delta_{\dot\alpha}}^{\dot\beta}{\delta_A}^B$. 
When contracting a vector field $V$ with a 
differential one-form $\alpha$ we use the notation $V\hook \alpha$.

With the capital Roman indices $A,B,\ldots $ ranging over both the bosonic
$\alpha,\beta,\ldots $ and the fermionic $i,j,\ldots$ indices
we use the  notation $ \{ AB \ldots ]$ for graded symmetrization 
and
$[AB\ldots \}$ for graded skew symmetrization
\begin{subequations}
\begin{eqnarray}
T_{\{A_1A_2\ldots A_n]}
\!\!&:=&\!\! \tfrac{1}{n!}\sum_{\sigma\in P_n}(-)^{\bar\sigma}T_{A_{\sigma
    (1)}A_{\sigma(2)}\ldots A_{\sigma(n)}},\\
T_{[A_1A_2\ldots A_n\}}
\!\!&:=&\!\! \tfrac{1}{n!}\sum_{\sigma\in P_n}(-)^{\bar\sigma+ |\sigma|}T_{A_{\sigma
    (1)}A_{\sigma(2)}\ldots A_{\sigma(n)}},
\end{eqnarray}
\end{subequations}
where $P_n$ is the group of permutations of $n$ letters, 
$|\sigma|$ the number of transpositions in $\sigma$ and
$\bar{\sigma}$ the number of transpositions of odd indices.

For an index such as $A$ that ranges over indices for both odd and
even coordinates, $p_A$ will denote the Gra{\ss}mann parity of the index, $p_A=0$
for an even coordinate, and 1 for an odd one so that a graded skew
form $\Lambda_{AB}$ satisfies
\begin{equation}
\Lambda_{AB}\ =\ -(-)^{p_Ap_B}\Lambda_{BA}. 
\end{equation}
We introduce $\epsilon_{\dot\alpha\dot\beta}=\epsilon_{[\dot\alpha\dot\beta]}$ with
$\epsilon_{\dot0\dot1}=-1$ and $\epsilon_{\dot\alpha\dot\gamma}\epsilon^{\dot\gamma\dot\beta}
={\delta_{\dot\alpha}}^{\dot\beta}$, 
and similarly for $\epsilon_{\alpha\beta}$.  

In the supersymmetric setting, there is a distinction between
differential and integral forms, the latter being required for
integration, Manin (1988).  Unless otherwise stated, all our forms will be
differential.

\subsection{Self-dual supergravity equations}
We introduce connections on $\cH$ and $\widetilde\cS$ represented by
connection one-forms ${\omega_A}^B$ and ${\omega_{\dot\alpha}}^{\dot\beta}$,
respectively.  These determine a connection $\nabla$ on $T\cM$ by
\begin{equation}
\nabla V^{A\dot\alpha}
   \ =\ \d V^{A\dot\alpha}+V^{B\dot\alpha}{\omega_B}^A
   +V^{A\dot\beta}{\omega_{\dot\beta}}^{\dot\alpha}
\end{equation}
so that it preserves the factorisation $T\cM\cong\cH\otimes\widetilde\cS$.
The fermionic parts of ${\omega_A}^B$ gauge the $R$-symmetry.

In this supersymmetric context, a choice of scale or volume form on
$\cM$ is a section of the Berezinian of $\Omega^1\cM$. We can assume
that the Berezinians of $\cH$ and $\widetilde \cS$ have been
identified so that the scale is determined by a section of the
Berezinian of either $\cH^*$ or $\widetilde \cS^*$.  The connections
can be chosen uniquely so that they preserves these sections of the
Berezinians of $\cH^*$ and $\widetilde\cS^*$ and so that the
connection on $T\cM$ has torsion with vanishing
supertrace.\footnote{Special care needs to be taken for $\CN=4$, Wolf
  (2007).}  We assume from hereon that such choices have been made.
In the formulae that follow, we will also assume that the connection
is torsion-free as that is part of the self-dual Einstein condition
(the torsion will not in general vanish on the full super space-time,
only on this right-chiral (or left-chiral) reduced supermanifold).

The curvature two-form ${R_{A\dot\alpha}}^{B\dot\beta}$ of
$\nabla$ decomposes into curvature two-forms for the connections on
$\cH$ and $\widetilde\cS$
\begin{equation}
 {R_{A\dot\alpha}}^{B\dot\beta}\ =\ {\delta_A}^B
{R_{\dot\alpha}}^{\dot\beta} + {\delta_{\dot\alpha}}^{\dot\beta}
{R_A}^B\, .
\end{equation}
 Making explicit the form indices, we write
the Ricci identities as
\begin{eqnarray}
   [\nabla_{A\dot\alpha},\nabla_{B\dot\beta}\}V^{D\dot\delta} \!\! &=&\!\!
   (-)^{p_C(p_A+p_B)}V^{C\dot\delta}
      {R_{A\dot\alpha B\dot\beta C}}^D +\notag \\
    &&\kern1cm+\ (-)^{p_D(p_A+p_B)}V^{D\dot\gamma}
    {R_{A\dot\alpha B\dot\beta\dot\gamma}}^{\dot\delta},
\end{eqnarray}
where $V^{A\dot\alpha}$ is a vector field on $\cM$.

In the torsion free case, using the algebraic Bianchi identities,
Prop. 2.6 of Wolf (2007) gives the decomposition of the
curvature into irreducibles:
\begin{subequations}\label{irred}
\begin{eqnarray}
R_{A\dot\alpha B\dot\beta C}{}^D\!\!&=&\!\!
-2(-)^{p_C(p_A+p_B)}R_{C[A|\dot\alpha\dot\beta|} \delta_{B\} }{}^D +
\epsilon_{\dot\alpha\dot\beta}{R_{ABC}}^D,\\ 
{R_{ABC}}^D\!\!&=&\!\!  C_{ABC}{}^D
-2(-)^{p_C(p_A+p_B)}\Lambda_{C\{ A}\delta_{B]}{}^D,\\
R_{A\dot\alpha B\dot\beta \dot\gamma}{}^{\dot\delta}\!\!&=&\!\!
C_{AB\dot\alpha\dot\beta\dot\gamma}{}^{\dot\delta} + 2
\Lambda_{AB}\delta_{(\dot\alpha}{}^{\dot\delta}\epsilon_{\dot\beta)\dot\gamma}
+ \epsilon_{\dot\alpha\dot\beta}R_{AB\dot\gamma}{}^{\dot\delta},
\end{eqnarray}
\end{subequations}
where 
the curvature tensors satisfy the
algebraic conditions
\begin{eqnarray}
R_{AB\dot\alpha\dot\beta}
\ =\ R_{AB\dot\alpha}{}^{\dot\gamma}\epsilon_{\dot\gamma\dot\beta}
\ =\ R_{AB(\dot\alpha\dot\beta)},\notag\\
{C_{ABC}}^D\ =\ {C_{\{ABC]}}^D\, ,
\quad (-)^{p_C}{C_{ABC}}^C\ =\ 0, \quad \Lambda_{AB}\ =\ \Lambda_{[AB\}}.
\end{eqnarray}  
Here, $\Lambda_{AB}$ is a natural supersymmetric extension of the
scalar curvature and will be set equal to the cosmological constant
when the field equations are satisfied.  (See Wolf (2007) for further
details of the construction and properties of the connections.)

\begin{defn}\label{sddef}
  A right-chiral superspace will be said to satisfy the $\CN$-extended
  self-dual supergravity equations if
\begin{itemize}
\item[(i)] the unique connection that preserves the given Berezinians
  of $\cH^*$ and $\widetilde\cS^*$ is torsion-free and satisfies
  $C_{AB\dot\alpha\dot\beta\dot\gamma}{}^{\dot\delta}=0$,
\item[(ii)]  $R_{AB\dot\alpha\dot\beta}=0$,
\item[(iii)] preserves some $P^{AB}=P^{[AB\}}\in
  \Lambda^2\cH$ of rank $2|r$ and is flat on the odd $(\CN-r)$-dimensional subspace of
  $\cH^*$ that annihilates $P^{AB}$.
\end{itemize}

When $\Lambda_{AB}\neq 0$ it is will be said to be Einstein, whereas
if $\Lambda_{AB}= 0$ it will be said to be vacuum.
When $r= 0$, the connection on $\cH$ is trivial in the odd directions and
  the $R$-symmetry is ungauged; all supersymmetry generators are
  covariantly constant.  For $r>0$, a subgroup of the
  $R$-symmetry is gauged with gauge group an extension of $SO(r,\IC)$, the
  subgroup of $SO(\CN,\IC)$ that preserves $P^{ij}$ the odd-odd part of
  $P^{AB}$.  For  $r=\CN$, the gauge group is $SO(\CN,\IC)$.  
\end{defn}
Conformal supergravity corresponds to the more general situation where
condition (i) alone is satisfied, and a natural supersymmetric
analogue of the hypercomplex
case corresponds to conditions (i) and (ii). In this work, we shall
mostly be concerned with the situation where (i)--(iii) are satisfied
simultaneously. 

There is only one possibility for the gauging in the Einstein case as follows:
\begin{lemma}
Either $P^{AB}$ and $\Lambda_{AB}$ both have maximal rank and can be chosen
to be multiples of each-other's inverse, or $\Lambda_{AB}=0$.
\end{lemma}

\proof
Condition (iii) of Def.~\ref{sddef} implies that
$(-)^{p_C+p_C(p_D+p_E)}{R_{ABC}}^{[D}P^{E\}C}=0$ and taking a 
supertrace gives the equation
\begin{equation}
0\ =\ (-)^{p_C(p_A + p_E)+p_C+p_B}\Lambda_{C\{ A}\delta_{B]}^{[B}P^{E\}C}
\end{equation}
which quickly leads to the condition that $(-)^{p_B}\Lambda _{AB}P^{BC}$ is a
multiple of ${\delta_A}^C$.  If this multiple is non-zero, $P^{AB}$ and
$\Lambda_{AB}$ have maximal rank and are multiples of each other's
inverse.  If this multiple is zero, the assumption on the rank of
$P^{AB}$ implies that the
rank of $\Lambda_{AB}$ is less than or equal to $0|\CN-r$.  The
condition that the connections is flat on the subspace of $\cH^*$ that
annihilates $P^{AB}$ implies that ${R_{ABC}}^De_D=0$ for all $e_D$
such that $P^{AB}e_B=0$.  Symmetrizing over $ABC$ gives that
${C_{ABC}}^De_D=0$ so we must also have
${\Lambda_{C\{A}}e_{D]}=0$. Note that for $\CN=2$ the multiple is always zero. 
$\Box$

\bigskip

It is a consequence of the Bianchi
identities that $\Lambda_{AB}$ is covariantly constant so that, when non-zero,
defining $P_{AB}$ as the inverse of $P^{AB}$, we can set
$\Lambda_{AB}=\Lambda P_{AB}$.
 When $\Lambda_{AB}$ is non-trivial,
 the curvature is non-trivial on the odd
 directions of $\cH$, and so the $R$-symmetry is therefore necessarily
 gauged with gauge group $SO(\CN,\IC)$.  

We will see in \S\ref{nonlinext} in the
 discussion of the deformations of twistor space how the different
 gaugings come about.

We also obtain 
$(-)^{p_C+p_C(p_D+p_E)}{C_{ABC}}^{[D}P^{E\}C}=0$ and
$\nabla_{[A\dot\alpha}{C_{B\}CD}}^E=0$. 
The field equations of self-dual supergravity with zero
cosmological constant lead to the Ricci identities
\begin{equation}
[\nabla_{A\dot\alpha},\nabla_{B\dot\beta}\}V^{D\dot\delta}\ =\
   (-)^{p_C(p_A+p_B)}V^{C\dot\delta}
\epsilon_{\dot\alpha\dot\beta}{C_{ABC}}^D
\end{equation}
which in turn imply Ricci-flatness of $\cM$.  

The self-dual supergravity equations on chiral super space-time with
vanishing cosmological constant first appeared in light-cone gauge and
in their covariant formulation in the work by Siegel
(1992).\footnote{See also Kallosh (1979,1980), Christensen, Deser,
   Duff \& Grisaru (1979), Ketov, Nishino \& Gates (1992) and
   Bergshoeff \& Sezgin (1992).}

\subsection{Twistor constructions}

Flat 
supertwistor space is $\IP\IT'_{[\CN]}:=\CP^{3|\CN}\setminus\CP^{1|\CN}$ with
homogeneous coordinates 
\begin{equation}
Z^I\ :=\ (\omega^{\alpha},\theta^i,\pi_{\dot\alpha})\ =\
(\omega^A,\pi_{\dot\alpha}),
\end{equation}
where $\omega^\alpha$ and $\pi_{\dot\alpha}$ are bosonic coordinates
and $\theta^i$ fermionic ones.

The supertwistor correspondence is between right-chiral complexified
super space-time $\IM_{[\CN]}\cong\IC^{4|2\CN}$ with coordinates
$(x^{\alpha\dot\alpha},\theta^{i\dot\alpha})=x^{A\dot\alpha}$, and is
expressed by the incidence relation
\begin{equation}
\omega^A\ =\ x^{A\dot\alpha}\pi_{\dot\alpha}\, .
\end{equation}
By holding $x^{A\dot\alpha}$ constant we see that points of
$\IM_{[\CN]}$ correspond to $\CP^1$s in supertwistor space
$\IP\IT'_{[\CN]}$ with homogeneous coordinates
$\pi_{\dot\alpha}$. Alternatively, by holding $Z^I$ constant, 
we see that points in $\IP\IT'_{[\CN]}$ correspond to $(2|\CN)$-dimensional
isotropic superplanes.  In the curved case, both sides of the
correspondence are deformed, but points of super space-time still
correspond to $\CP^1$s in supertwistor space (and points of
supertwistor space to $(2|\CN)$-dimensional isotropic
subsupermanifolds of $\cM$).

Bosonic twistor space will be denoted by $PT$ and will be a
deformation of some region in $\CP^3$, whereas a supersymmetrically
extended curved twistor space will be denoted by $\cP\cT$ and will be
a deformation of a region in $\CP^{3|\CN}$. Similarly, a bosonic
space-time will be denoted by $M$ and a supersymmetric one (which will
always in this paper be right-chiral) by $\cM$.

We recall first Ward's (1980) extension of Penrose's (1976) non-linear
graviton construction to the case of non-zero cosmological constant:

\begin{thm}{\bf (Penrose 1976, Ward 1980)}
 \begin{itemize}
 \item[(i)] There is a natural one-to-one correspondence between
   holomorphic conformal structures $[g]$ on some four-dimensional
   (complex) manifold $M$ whose anti-self-dual Weyl curvature
   vanishes, and three-dimensional complex manifolds $PT$ (the twistor
   space) containing a rational curve (a $\CP^1$) with normal bundle
   $N\cong\cO(1)\oplus \cO(1)$.
  \item[(ii)]
  The existence of a conformal scale for which the trace-free Ricci
  tensor vanishes, but for which the scalar curvature is
  non-vanishing, is equivalent to $PT$ admitting a non-degenerate
  contact structure.
\item[(iii)] The existence of a conformal scale for which the full
  Ricci tensor vanishes is equivalent to $PT$ admitting a fibration
  $\varpi:PT\rightarrow \CP^1$ whose fibres admit a Poisson structure
  with values in the pullback of $\cO(-2)$ from $\CP^1$.
\end{itemize}
\end{thm}

\noindent
Here, $\cO(n)$ is the complex line bundle of Chern class $n$ on
$\CP^1$.

The holomorphic contact structure is a rank-$2$
distribution $D \subset T^{(1,0)}PT$ in the holomorphic tangent bundle
of $PT$.  The quotient determines a line bundle $L:=T^{(1,0)}PT/D$.
It can be defined dually to be the kernel of a holomorphic
$(1,0)$-form $\tau$ defined up to scale on $PT$, i.e. $D=\ker\tau$.
If so, $\tau$ takes values in $L$ since the map
$T^{(1,0)}PT\rightarrow T^{(1,0)}PT/D:=L$ is then the contraction of
a vector with the $(1,0)$-form $\tau$. The non-degeneracy condition is
that for any two vector fields $X$ and $Y$ in $D$, the Frobenius form
\begin{equation}
\Phi\,:\,D\wedge D\
\to\ L\ :=\ T^{(1,0)}PT/D,\quad\mbox{with}\quad
  \Phi(X,Y)\ :=\  [X,Y] \mbox{ mod } D
\end{equation}
is non-degenerate on $D$.  This is equivalent to $\tau\wedge\d\tau\neq
0$.  When it is everywhere degenerate, $D$ determines a foliation whose leaves
are the fibres of the projection $\varpi:PT\rightarrow \CP^1$ and $\tau$ is
the pullback of the one-form $\pi^{\dot\alpha}\d \pi_{\dot\alpha}$ from
$\CP^1$ and $L$ becomes the pullback of $\cO(2)$ from $\CP^1$.  In the
non-degenerate case, we can define a Poisson structure with values in
$L^{*}$ to be the inverse of $\Phi$ on $D$.  This has an analogue
also in the degenerate case, now with values in $\cO(-2)$ although its
existence no longer follows from that of $\tau$.

We can impose compatibility with, e.g., Euclidean reality conditions
by requiring the existence of an anti-holomorphic involution
$\rho\,:\,PT\to PT$ without fixed points sending the given Riemann
sphere to itself via the antipodal map.  This then induces a
corresponding involution on $M$ fixing a real slice on which
the metric $g$ is
real and of Euclidean signature.

The above theorem has a supersymmetric extension as follows:
\begin{thm} \label{wardthm}
{}
\begin{itemize}
 \item[(i)] 
  There is a natural one-to-one correspondence between conformally
  self-dual holomorphic right-chiral space-times and complex
  supermanifolds $\cP\cT$ of dimension $3|\CN$ with an embedded
  rational curve (a Riemann sphere $\CP^1$) with normal bundle
  $\cN\cong\cO(1)^{\oplus 2|\CN}$.
 \item[(ii)]
  Furthermore, $\cM$ is a complex solution to the four-dimensional
  $\CN$-extended self-dual super\-gravity equation with
  non-vanishing cosmological constant iff the twistor space $\cP\cT$
  admits a non-degenerate even contact structure.
 \item[(iii)]
  $\cM$ is a complex solution to the four-dimensional
  $\CN$-extended self-dual super\-gravity equation with
  vanishing cosmological constant iff the twistor space $\cP\cT$
  admits a fibration $\varpi:\cP\cT\rightarrow \CP^{1|\CN-r}$ and a Poisson
  structure of rank $2|r$ tangent ot the fibres with values in $\varpi^*\cO(-2)$.
\end{itemize}
\end{thm}

\noindent
Here, $\cO(n)^{\oplus r|s}:=\mathbbm{C}^{r|s}\otimes \cO(n)$.  The
proof breaks up into three parts; further details
of the non-degenerate cosmological constant case are given in Wolf (2007).

\proof
Part (i). Let $\cF=\mathbbm{P}(\widetilde\cS^*)$ be the projective co-spin
bundle over $\cM$
with holomorphic projection $p\,:\,\cF\to\cM$. Its fibres
$p^{-1}(x)$ over $x\in\cM$ are complex projective lines
$\CP^1$ with homogeneous fibre coordinates
$\pi_{\dot\alpha}$.  We define the twistor distribution to be the
rank-$2|\CN$ distribution $\cD_\cF$ on $\cF$ given by
\begin{equation}\label{twistor-dist}
 \cD_\cF\ :=\ {\rm span}\{\widetilde E_A\}\ :=\
{\rm span}\left\{\pi^{\dot\alpha} E_{A\dot\alpha} +
    \pi^{\dot\alpha}\pi_{\dot\gamma}
    {\omega_{A\dot\alpha\dot\beta}}^{\dot\gamma}
    \frac{\p}{\p\pi_{\dot\beta}}\right\}\!,
\end{equation}
where the $E_{A\dot\alpha}$s are the frame fields and
${\omega_{\dot\alpha}}^{\dot\beta}$ is the connection one-form on
$\widetilde\cS$. A few lines of algebra show that $\cD_\cF$ is
integrable if and only if the connection is torsion-free and the
$C_{AB(\dot\alpha \dot\beta\dot\gamma\dot\delta)}$-part of the
curvature vanishes. In this case, the distribution $\cD_\cF$
defines a foliation of $\cF$. Working locally on $\cM$, the resulting
quotient will be our supertwistor space, a $(3|\CN)$-dimensional
supermanifold denoted by $\cP\cT$.  The quotient map will be denoted
by $q\,:\,\cF\to\cP\cT$ so that we have the double fibration
$\cP\cT\stackrel{q}\leftarrow \cF\stackrel{p}\rightarrow\cM$.  We note
that we can form a non-projective supertwistor space $\cT$ by taking
the quotient of $\widetilde \cS^*$ by the distribution $\cD_\cF$.  The
integral curves of the Euler vector field
$\widetilde\Upsilon:=\pi_{\dot\alpha}\p/\p\pi_{\dot\alpha}$ are the
fibres over $\mathbbm{P}(\widetilde\cS^*)$ and $\widetilde \Upsilon$
descends to give a vector field $\Upsilon$ on $\cT$ which determines
the fibration $\cT\rightarrow\cP\cT$.

Since $\cF$ is a $\CP^1$-bundle over $\cM$ and the fibres are
transverse to the distribution $\cD_\cF$, the submanifolds
$q(p^{-1}(x))\hookrightarrow\cP\cT$, for $x\in\cM$, are
$\CP^1$s. In the other direction, the supermanifolds
$p(q^{-1}(Z))\hookrightarrow\cM$, for $Z\in\cP\cT$, are the
$(2|\CN)$-dimensional isotropic subsupermanifolds of $\cM$ given by
the $p$ projections of integral surfaces of $\cD_\cF$.

The inverse construction, i.e.~starting from $\cP\cT$, follows by
applying a supersymmetric extension of Kodaira's deformation theory
(Waintrob 1986).  This allows one to reconstruct $\cM$ as the moduli
space of $\CP^1$s that arise as deformations of the given $\CP^1$
which will correspond to some $x\in\cM$.  According to Kodaira theory,
$T_x\cM\cong H^0(\CP^1,\cN)$, where $\cN$ is the normal bundle to the given
$\CP^1\subset\cP\cT$, and in order that the moduli space exist, we
require the vanishing of the first cohomology of the normal bundle
$\cN$.  If the given $\CP^1$ arises as $q(p^{-1}(x))$ for some
$x\in\cM$, then $\cN\cong\cO(1)^{\oplus 2|\CN}$: this can be seen by
expressing it as the quotient of the horizontal tangent vectors to
$\cF$ at $p^{-1}(x)\cong\CP^1$, which can be represented by
$T_x\cM$, by $\cD_\cF$
\begin{equation}
 0\ \longrightarrow\ \cD_\cF|_{p^{-1}(x)} \longrightarrow\
T_x\cM \
\longrightarrow\    q^*\cN\ \longrightarrow\ 0\, . 
\end{equation}
Since the twistor distribution $\cD_\cF$ restricted to the fibres
$p^{-1}(x)$ over $x\in\cM$ is $\cO(-1)^{\oplus 2|\CN}$, and
$T_x\cM\cong\IC^{4|2\CN}$,
$\cN$ takes the form $\cO(1)^{\oplus 2|\CN}$ as stated above. Kodaira
theory in turn implies that we can reconstruct $\cM$ as the moduli
space of such $\CP^1$s, and that the construction is stable
under deformations of the complex structure on $\cP\cT$.  Kodaira
theory identifies the tangent bundle $T_x\cM$ with the sections of the
normal bundle,   $\cN\cong\cO(1)^{\oplus 2|\CN}$, and these, by an
extension of Liouville's theorem are linear functions of
$\pi_{\dot\alpha}$, i.e., $V^{A\dot\alpha}\pi_{\dot\alpha}$ where the
  $A$ index is associated to a basis of $\mathbbm{C}^{2|\CN}$.
This gives the right-chiral manifold structure on $\cM$, and it is
easily seen that lines through a given point of $\cP\cT$ correspond to
an integrable $(2|\CN)$ manifold that will be an integral surface of
the distribution $\cD_\cF$.  Thus $\cD_\cF$ is integrable and the
$\cM$ is therefore conformally self-dual.

\smallskip

Part (ii). In the self-dual Einstein case with non-vanishing cosmological
constant, we may introduce a  one-form of
homogeneity $2$ on $\cF$ by
\begin{equation}
\widetilde\tau\ :=\ \pi^{\dot\alpha}\nabla\pi_{\dot\alpha}\
=\ \pi^{\dot\alpha}\d\pi_{\dot\alpha}-
{\omega_{\dot\alpha}}^{\dot\beta}\pi^{\dot\alpha}\pi_{\dot\beta},
\end{equation}
where ${\omega_{\dot\alpha}}^{\dot\beta}$ is the connection one-form
on $\widetilde\cS$. The one-form $\tau$ automatically annihilates
horizontal vectors and hence the distribution $\cD_\cF$.  The form
$\widetilde\tau$ descends to $\cP\cT$ if and only if $\d
\widetilde\tau$ is annihilated by $\cD_\cF$ also.  This characterizes
the self-dual Einstein equations since  when
$C_{AB\dot\alpha\dot\beta\dot\alpha\dot\delta}=0$, as follows form the
conformal self-duality condition,
\begin{equation}
 \d\widetilde\tau\ =\
\nabla\pi^{\dot\alpha}\wedge\nabla\pi_{\dot\alpha} +
E^{B\dot\beta}\wedge E^{A\dot\alpha}\Lambda_{AB}
\pi_{\dot\alpha}\pi_{\dot\beta} - E^{B\dot\gamma}\wedge E^A_{\dot\gamma}R_{AB\dot\alpha\dot\beta} \pi^{\dot\alpha}\pi^{\dot\beta}\, .
\end{equation}
and this is annihilated by $\cD_\cF$ iff
$R_{AB\dot\alpha\dot\beta}=0$.  Thus, $\widetilde\tau$ descends to
$\cP\cT$, i.e., there exists a one-form $\tau$ on $\cP\cT$ such that
$\widetilde\tau=q^*\tau$.

Non-degeneracy of the contact structure is the condition that $\d\tau$
is non-de\-generate on the kernel $\cD$ of $\tau$, or equivalently,
the condition that the three-form $\tau\wedge\d\tau$ should be
non-degenerate in the sense that for any vector $X$,
$X\hook(\tau\wedge\d\tau)=0 \Rightarrow X=0$. This
non-degeneracy is equivalent to the non-degeneracy of $\Lambda_{AB}$
on $\cH$.  Thus, $\tau$ defines a non-degenerate holomorphic contact
structure on $\cP\cT$.

\smallskip

Part (iii). In the self-dual vacuum case, we see that the connection on
$\widetilde\cS$ is flat and a basis for $\widetilde \cS$ can be found
so that it vanishes.  In this basis, $\pi_{\dot\alpha}$ are constant
along the horizontal distribution on $\cF$, and so along the
distribution \eqref{twistor-dist}.  They are therefore the pullback of
coordinates on $\cP\cT$. The condition that the connection is flat on the
annihilator of $P^{AB}$ in $\cH^*$ means that there are $\CN-r$
covariantly constant  sections $e^s_{A}$
of the odd part of $\cH^*$,  $s=r+1,\ldots ,\CN$.  The  forms
$E^{A\dot\alpha}e^s_{A}$ are therefore constant and, since the
connection is torsion free, these forms are exact and equal to
$\d\theta^{s\dot\alpha}$ for some odd coordinates
$\theta^{s\dot\alpha}$.  The $\CN-r$ functions
$\theta^s=\theta^{s\dot\alpha}\pi_{\dot\alpha}$ can be seen to be
constant also along the twistor distribution \eqref{twistor-dist}.
The global holomorphic coordinates $(\pi_{\dot\alpha},\theta^s)$ define
a projection $\varpi:\cP\cT\rightarrow \CP^{1|\CN-r}$ as promised.
We now define the Poisson structure by considering a pair of local
functions $f,g $ on $\cP\cT$.  Pulled back to $\cF$, they satisfy
\begin{equation}
\pi^{\dot\alpha}E_{A\dot\alpha} f\ =\ 0, \quad
\pi^{\dot\alpha}E_{A\dot\alpha} g\ =\ 0
\end{equation}
and this implies that
\begin{equation}
E_{A\dot\alpha}f\ =\ \pi_{\dot\alpha} f_A,\quad
E_{A\dot\alpha}g\ =\ \pi_{\dot\alpha} g_A 
\end{equation}
for some $f_A$, $g_A$ of weight $-1$ in $\pi_{\dot\alpha}$ (this
follows from the standard fact that
$\pi^{\dot\alpha}b_{\dot\alpha}=0\Rightarrow
b_{\dot\alpha}=b\pi_{\dot\alpha}$ for some $b$ which follows from the
two-dimensionality of the spin space and the skew symmetry of
$\epsilon_{\dot\alpha\dot\beta}$).  We define the Poisson bracket
$\{f,g\}$ of $f$ with $g$ to be
\begin{equation}
\{f,g\}\ :=\ (-)^{p_A(p_f+1)}f_A P^{AB}g_B.
\end{equation}
It is clear that this has weight $-2$ in $\pi_{\dot\alpha}$, but, as
given, this expression only lives on $\cF$.  However, it is easily
checked that, as a consequence of the covariant constancy of the
$P^{AB}$, it is constant along the distribution \eqref{twistor-dist}
and descends to $\cP\cT$.  \hfill$\Box$

\bigskip 

%For $\CN$ even, the Einstein case can also be characterized in terms
%of a Poisson structure defined by
%\begin{equation}
%\{f,g\}\tau\wedge(\d\tau)^{\wedge (1+\CN/2)}
%\ =\ \d f\wedge \d g \wedge \tau\wedge(\d\tau)^{\wedge \CN/2}.
%\end{equation}
See appendix \ref{nonproj} for more on the non-projective formulation.

\section{Twistor actions}\label{twistoraction}

In order to consider actions, we must allow our fields to go
off-shell, and this is most straightforwardly done in the Dolbeault
setting.  We can take an almost complex structure that is not
necessarily integrable to be the off-shell field, and regard the
integrability condition to be part of the field equations.  In the
following we will see that if we require the almost complex
structure to be compatible with a Poisson structure or complex contact
structure and the almost complex structure can be encoded in a
complex one-form $h$ defined up to scale.

In the following, we will mostly work `non-projectively' i.e., on
$\IT_{[\CN]}=\IC^{4|\CN}$, or at least using homogeneous coordinates.
This can also be identified as the total space of the line bundle
$\cO(-1)$ over $\IP\IT$.  On this space, we have the Euler homogeneity
vector field $\Upsilon$, and a canonically defined holomorphic volume
form $\Omega$ (an integral form in this supersymmetric context) of
weight $4-\CN$, the tautological form pulled back from ${\rm
  Ber}(\cP\cT)\cong\cO(\CN-4)$ satisfying ${\mathcal L}_\Upsilon
\Omega=(4-\CN)\Omega$, where $\mathcal{L}_\Upsilon$ is the Lie
derivative along $\Upsilon$.  Similarly, $\tau$ will be a well-defined
differential one-form of weight 2.  See appendix \ref{nonproj} for
further discussion.

\subsection{Deformations of twistor space}\label{nonlinext}

For simplicity, we take the supertwistor space $\cP\cT$ to be a 
deformation of flat twistor space $\IP\IT'_{[\CN]}$ with homogeneous
coordinates as in the flat case given by\footnote{We could take a
  finite deformation of any curved integrable twistor space, but would
  then need more coordinate patches.}
\begin{equation}
 Z^I\ =\ (\omega^\alpha,\theta^i,\pi_{\dot\alpha})\ =\ 
 (\omega^A,\pi_{\dot\alpha})\ =\ (Z^a,\theta^i)
\end{equation}
the latter form distinguishes between the odd, $\theta^i$ and the
even, $Z^a$ coordinates. We also assume that we are given an `infinity
twistor' a constant graded skew bi-vector 
\begin{subequations}
\begin{equation}
I^{IJ}\ :=\ \mathrm{diag} (P^{AB},\Lambda
\epsilon_{\dot\alpha\dot\beta}),
\end{equation}
where
\begin{equation}
P^{AB}\ =\ \mathrm{diag} (\epsilon^{\alpha\beta},
P^{ij})\qquad\mbox{and}\qquad P^{ij}\ =\ P^{(ij)}.
\end{equation}
\end{subequations}
When $\Lambda=0$, we will take $P^{ij}$ to be diagonal with $r$ ones and
$\CN-r$ zeroes along the diagonal.

We also introduce the graded Poisson structure on homogeneous
functions $f$ and $g$ by
\begin{subequations}
\begin{equation}
[f,g\}\ :=\ (-)^{p_I(p_f+1)}(\p_If) I^{IJ}(\p_Jg),
\end{equation}
where we introduce the notation 
\begin{equation}
 \p_I\ :=\ \frac{\p}{\p Z^I}, \qquad \mbox{ and we will also use } \qquad \bar\p_{\bar
  I}\ :=\ \frac{\p}{\p \bar Z^{\bar I}}.
\end{equation}
\end{subequations}

Infinitesimally, a deformation of the almost complex structure is
represented by a holomorphic tangent bundle valued $(0,1)$-form $j$,
where the deformed and undeformed anti-holomor\-phic exterior
derivatives are related by $\bar\p=\bar\p_0 + j$.  The first order
part of the integrability condition (assuming that $\bar\p_0^2=0$) is
$\bar\p_0 j=0$.  An infinitesimal diffeomorphism induced by the real
part of a $(1,0)$-vector field $X$ gives rise to the deformation
$j:=-\bar\p_0 X$, so that the infinitesimal deformations of the
complex structure modulo those obtained by infinitesimal
diffeomorphisms define an element of the Dolbeault cohomology group
$H^1(\IP\IT',T^{(1,0)}\IP\IT)$.

In order to impose the Einstein or vacuum conditions, we will also demand that
the deformation preserves the Poisson structure $\Pi=-I^{JI}\p_I\wedge\p_J$ of
weight $-2$.  In
this linearised context, we can ensure this by requiring that the
deforming vector fields $j$ preserve the Poisson structure 
$\CL_j \Pi=0$, where $\CL$ is the Lie derivative.
This will follow if $j$ is Hamiltonian with respect to $\Pi$, i.e., if
there exists a $(0,1)$-form $h$ of weight 2 such that 
\begin{equation}
j\ =\ \Pi\hook \d h\ =\  (-)^{p_I}(\p_Ih)I^{IJ}\p_J.
\end{equation}
If $h=\bar\p\chi$ we see that $j$ is
$\bar \p (\Pi(\chi))$ and so is pure gauge.  Thus such deformations
correspond to $h$ taken to be Dolbeault representatives for elements of
$H^1(\IP\IT',\cO(2))$.  The Penrose transform gives the identification
between elements of $H^1(\IP\IT',\cO(2))$ and linearised self-dual
gravitational fields, Penrose (1968, 1976) and in the supersymmetric case
this will give the whole associated linearised gravitational supermultiplet.

We now consider a {\it finite} deformation, again determined by
$h=\d\bar Z^{\bar a}h_{\bar a}$ which, at this stage, is an arbitrary
(even) smooth function of $(Z^I, \bar Z^{\bar a})$ homogeneous of
degree $2$ in $Z^I$ and 0 in $\bar Z^{\bar I}$, holomorphic in the
$\theta^i$s and satisfies $\bar Z^{\bar a}h_{\bar a}=0$; we will never
allow any dependence on the complex conjugates of the fermionic
cooordinates.

We then define the distribution $T^{(0,1)}\cP\cT$ of anti-holomorphic
tangent vectors on $\cP\cT$ by
\begin{equation}\label{eq:antiholoT}
T^{(0,1)}\cP\cT\ :=\ {\rm span}\{ \bar D_{\bar I}\}\ :=\
    {\rm span}\left\{\bar\p_{\bar a}+(-)^{p_I}(\p_Ih_{\bar a})I^{IJ}\p_J,
    \bar\p_{\bar i}\right\}\!.
\end{equation}
This is to be understood as a finite perturbation of the standard complex
structure on flat supertwistor space with $\bar\p$-operator $\bar\p_0=\d  \bar
Z^{\bar I}\bar\p_{\bar I}$.\footnote{As in the linearised context, we 
eventually want to impose the Einstein condition on the space-time 
manifold. Therefore, we are only
interested in a subclass of (finite) deformations $\bar\p_0\mapsto\bar\p_0+j$
with $j$ given by $j=\d\bar Z^{\bar a} {j_{\bar a}}^I\p_I=
\d\bar Z^{\bar a}(-)^{p_I}\p_I h_{\bar a}I^{IJ}\p_J$.}
The complex structure can be equivalently determined by specifying the
space of (1,0)-forms 
\begin{equation}\label{eq:holoforms}
\Omega^{(1,0)}\cP\cT\ :=\ {\rm span}\{ D Z^I\}\ :=\ {\rm span}\{\d Z^I +I^{IJ}\p_Jh\}.
\end{equation}

The integrability condition for this distribution is
\begin{equation}\label{maineq}
 I^{IJ}\p_J\left ( \bar\p_{\bar a} h_{\bar b}-\bar\p_{\bar b} h_{\bar a}
   + [h_{\bar a} \, ,h_{\bar b}\} \right)\ =\ 0~~\Longleftrightarrow~~
    I^{IJ}\p_J\left(\bar\p_0h+{\textstyle\frac 12}[h,h\}\right)\ =\ 0,
\end{equation}
where the wedge product in the last expression is understood.  When
this equation is satisfied, not only is the almost complex structure
integrable, but also the Poisson bracket of two holomorphic
functions is again holomorphic.  In the case that $\Lambda=0$, when
the Poisson structure is degenerate, the coordinates
$\pi_{\dot\alpha}$ and $\theta^{r+1},\ldots, \theta^\CN$ are holomorphic
and define a projection to $\CP^{1|\CN-r}$ as required for the
  characterization of a twistor space for a self-dual vacuum
  solution.  Thus, in this case, equation \eqref{maineq} is the main
  field equation.

In the Einstein case, we must produce a holomorphic contact structure.
On the flat twistor space, introduce the contact structure 
\begin{subequations}
\begin{equation}
\tau_0\ =\ \d Z^I Z^JI_{JI},
\end{equation}
where
\begin{equation}
 (-)^{p_K}I_{IK}I^{KJ}\ =\ \Lambda
{\delta_I}^J\qquad\mbox{and}\qquad I_{IJ}\ =\ \mathrm{diag} (\Lambda P_{AB},
\epsilon^{\dot\alpha\dot\beta}).
\end{equation}
\end{subequations}
For the Einstein case, from Thm.~\ref{wardthm}, we need to  
know that we have a holomorphic contact structure on the deformed space.  The
deformed one can be taken to be
\begin{equation}
\tau\ :=\ DZ^I Z^JI_{JI}\ =\ \d Z^I Z^J\omega_{JI}+
        Z^J\underbrace{(-)^{p_I}I_{JI} I^{IK}}_{=\ \Lambda{\delta_J}^K}\p_K
         h\ =\ \tau_0+2 \Lambda h,
\end{equation}
where the last equation follows from the homogeneity relation
$Z^I\p_Ih=2h$.  The condition that $\bar\p\tau=0$ $\Leftrightarrow$
$\bar D_{\bar I}\hook\d\tau=0$ is
\begin{equation}\label{eq:fieldeq}
F^{(0,2)}\ :=\ \bar\p_0h+{\textstyle\frac12}[h,h\}\ =\ 0.
\end{equation}
Thus, integrability of the complex structure follows from the holomorphy
of the contact structure when $\Lambda\neq 0$.  (When $\Lambda=0$,
$\tau_0$ remains holomorphic trivially.)  Thus, not only is 
\eqref{eq:fieldeq} our main equation in the Einstein case, it also
implies \eqref{maineq} in the other cases, and so we will focus on
this  as the main equation in what follows.

The choice of the Poisson structure reduces the diffeomorphism freedom to 
(infinitesimal) Hamiltonian coordinate 
transformations of the form
\begin{equation}\label{eq:gauge}
\delta Z^I\ =\ [Z^I,\chi\} \quad \leadsto \quad  h\ \mapsto\ h+\delta
h,\quad\mbox{with}\quad 
\delta h\ =\ \bar\p_0\chi+[h,\chi\},
\end{equation}
where $\chi$ is some smooth function of weight $2$.  Under this
transformation, 
the `curvature' $F^{(0,2)}$ behaves as $F^{(0,2)}\mapsto F^{(0,2)}
+\delta F^{(0,2)}$ with $\delta F^{(0,2)}=[F^{(0,2)},\chi\}$.
Thus, the field equation \eqref{eq:fieldeq} is invariant under these
transformations. 

We can see that, at least in linear theory, $h$ encodes a supergravity
multiplet as follows.  The form $h$ may be expanded in the odd
coordinates as
\begin{equation}
 h\ =\ h_0+\sum_{r=1}^\CN{\textstyle\frac{1}{r!}}\,\theta^{i_1}\cdots\theta^{i_r} h_{i_1\cdots i_r}.
\end{equation}
If we further linearise \eqref{eq:fieldeq} around the trivial solution
$h=0$, it tells us that $\bar\p_0 h=0$, or equivalently, $\bar\p_0
h_0=0=\bar\p_0 h_{i_1\cdots i_r}$. Because of the gauge invariance
\eqref{eq:gauge}, which at linearised level reduces to $\delta
h=\bar\p_0\chi$, we see that $h_0\in H^1(PT,\cO(2))$ and $h_{i_1\cdots
  i_r}\in H^1(PT,\cO(2-r))$, where $PT$ represents the body of the
supermanifold $\cP\cT$ (so that $PT$ is a finite deformation of
$\IP\IT'_{[0]}$.)  By virtue of the Penrose transform, Penrose (1968),
$h_0$ corresponds on space-time to a helicity $s=2$ field while
$h_{i_1\cdots i_r}$ to a helicity $s=(4-r)/2$ field.  Hence, for
maximal $\CN=8$ supersymmetry, we find
$(s_m)=(-2_1,-\frac32\,\!_8,-1_{28},
-\frac12\,\!_{56},0_{70},\frac12\,\!_{56},1_{28},\frac32\,\!_8,2_1)$
which is precisely the (on-shell) spectrum of $\CN=8$ Einstein
supergravity; the subscript `$m$' refers to the respective
multiplicity. Altogether, we see that a single element $h\in
H^1(\IP\IT,\cO(2))$ encodes the full particle content of maximally
supersymmetric linearised Einstein gravity in four dimensions.

In this linearised context, it is straightforward to see how the
gauging works.  The bundle of $R$-symmetry generators on twistor space
is the tangent bundle to the odd directions spanned by
$\p/\p\theta^i$.  The linearised variation in the $\bar\p$-operator on
this bundle is $P^{ik}\p^2h/\p\theta^j\p\theta^k $ because the part of
$\bar\p f^i\p/\p\theta^i$ tangent to the odd directions is $(\bar\p
f^i +P^{ik}\p^2h/\p\theta^j\p\theta^k f_j)\p/\p\theta^i$.  Because
$\theta^i$ anti-commute, $\p^2 h/\p\theta^i\p\theta^j$ is skew
symmetric in $ij$.  Thus, in the case of non-degenerate $P^{ij}$, this
gives an element of the Lie algebra of $SO(\CN,\IC)$, and so
corresponds to the maximal gauging of the $R$-symmetry, with gauge
group $SO(\CN,\IC)$.  When $P^{ij}$ has rank $r$, for $r<\CN$, the
gauging of the $R$-symmetry will be reduced to the subgroup of
$SO(\CN,\IC)$ that preserves $P^{ij}$.

In the appendix \ref{sec:app-pf}, where  we compare our
approach with that of Karnas \& Ketov (1998), we also make
some comments on the space-time fields in the non-linearised setting for zero
cosmological constant.

\subsection{Action functionals}
We will be interested in integrating Lagrangian
densities over twistor space for which we will need the holomorphic
volume {\it integral} form 
\begin{equation}
\Omega_\CN\ =\ D(D Z^I)\ =\ {\textstyle\frac{1}{4!}}\epsilon_{abcd}Z^a D Z^b\wedge
D Z^c \wedge D Z^d \otimes \prod_{i=1}^\CN D\theta^i
\end{equation}
which has weight $4-\CN$ on account of the Berezinian integration
rule $\int\d \theta^i \theta^j=\delta^{ij} $ implying $\d (\lambda \theta^i)
=\lambda^{-1}\d \theta^i$ for $\lambda\in\mathbbm{C}^*$. Here, we use
Manin's (1988) notation to denote integral forms associated
with a given basis of differential one-forms.  We will not integrate
over any complex conjugated odd coordinates.

For maximal supersymmetry, $\CN=8$,
we can write down an action functional reproducing the field equations
\eqref{eq:fieldeq} and hence also \eqref{maineq}
\begin{eqnarray}\label{eq:csaction}
S[h] \!\!&=&\!\! \int\Omega_8\wedge \left(h\wedge\bar\p_0 h + {\textstyle\frac 13}
h\wedge[h,h\}\right)\notag\\
    \!\!&=&\!\! \int\Omega_8^{(0)}\wedge \left(h\wedge\bar\p_0 h + {\textstyle\frac 13}
h\wedge[h,h\}\right)\!,
\end{eqnarray}
where the integral form
\begin{equation}
\Omega_8^{(0)}\ =\ D(\d Z^I)\ =\ {\textstyle\frac{1}{4!}}\epsilon_{abcd}Z^a \d Z^b\wedge
\d Z^c \wedge \d Z^d \otimes \prod_{i=1}^8 \d \theta^i.
\end{equation}
It can be seen that the weights balance as $h$ has weight 2,
$[\cdot,\cdot\}$ weight $-2$ and $\Omega_8$ (respectively,
$\Omega_8^{(0)}$) has weight $-4$. This is the only value of $\CN$ for
which there is such a balance. 

The action \eqref{eq:csaction} is invariant under \eqref{eq:gauge}. This
follows from the Bianchi identity for $F^{(0,2)}$, 
\begin{equation}\label{eq:bianchi}
\bar\p_0 F^{(0,2)}+[h,F^{(0,2)}\}\ =\ 0,
\end{equation}
implied by the (graded) Jacobi identity for the Poisson structure.

It is clear that the almost complex structure, integrability
conditions and action formulation (the latter for $\CN=8$) only depend
on the Poisson structure $ I^{IJ}$ and not on $I_{IJ}$ directly.  It
is also clear that if $ I^{IJ}$ is degenerate, the above field
equations and action (the latter for $\CN=8$) all make good sense,
although the action most directly yields \eqref{eq:fieldeq} rather
than the superficially weaker equation \eqref{maineq} that is
sufficient to determine the relevant structures on the deformed
twistor space.

The action \eqref{eq:csaction} can be compared
with the Kodaira-Spencer actions introduced in Bershadsky et
al.~(1994), the compendium of topological M-theory related actions
in Dijkgraaf et al.~(2005) and the Lagrange
  multiplier-type action involving the Nijenhuis tensor given
  in Berkovits \& Witten (2004) in the $\CN=4$ case.  Our action is
  local in contra-distinction with the non-local Kodaira-Spencer action. 
Our action is given
for a non-Calabi Yau space (due the isomorphism \eqref{eq:ber}, the
holomorphic Berezinian is only trivial when $\CN=4$).
Ours is most closely related to that in 
Berkovits \& Witten, although our basic
variable, the one-form $h$ which is a ``potential'' for
the deformation $j$, considered in deformation theory (i.e.~$j$ is a
holomorphic derivative of $h$) and is most naturally expressed for
$\CN=8$ rather than $\CN=4$.

We close this subsection by discussing the cases with $\CN < 8$
supersymmetries. We start from the action \eqref{eq:csaction} with
$\CN=8$ but
restrict the dependence of $h$ on $\theta^i$ by requiring invariance
under an $SO(8-\CN,\IC)$ subgroup of the $R$-symmetry.  Thus, we set
\begin{equation}
h\ =\ f+\theta^{\CN+1}\cdots \theta^8\, b,
\end{equation}
where $f$ and $b$ are now one forms depending on the bosonic twistor
coordinates and $\theta^1, \ldots, \theta^\CN$, $f$ has weight 2, and $b$
has weight $\CN-6$.  We can now integrate out
the anti-commuting variables $ \theta^{\CN+1}, \ldots, \theta^8$ 
and integrate by parts to obtain the action
\begin{equation}\label{eq:Bfaction}
S[b,f]\ =\ \int \Omega_r \wedge b\wedge \left(  \bar\p_0 f + \tfrac{1}{2}
  [f,f\}\right)\!.
\end{equation}
This action is now of `BF' form where $b$ acts as a Lagrange
multiplier for the field equation
\begin{equation}
 \bar\p_0 f + \tfrac{1}{2}[f,f\}\ =\ 0.
\end{equation}
which, as we have seen, implies that integrability of the complex
structure compatible with a holomorphic Poisson structure.  Varying
$f$ yields the equation
\begin{equation}
\bar\p _f b\ =\ 0
\end{equation}
and, together with the gauge freedom $b\mapsto b+\bar\p_f \chi$,
this implies that $b$ defines an element of the cohomology group
$H^1(\cP\cT, \cO(\CN-6))$ and so is the Penrose transform of a
superfield of helicity $-2 +\CN/2$.

\section{Covariant approach, covariant action for $\CN=0$ and
special geometry}

The above actions are non-covariant in the sense that they explicitly
depend on the chosen background one has started with so that
diffeomorphism invariance is broken. This is normal in the context of
Chern-Simons actions for which a frame of the Yang-Mills bundle must
be chosen.  Nevertheless, we will see
that at least for $\tau$ non-degenerate and $\CN=0$ we can give a
covariant version. 

The geometric structure we are concerned with here is closely related
to a (real) six-dimensional special geometry introduced by Cap \&
Eastwood (2003).  In their geometry, a real rank-$4$ distribution
(subbundle of the tangent bundle) $D$ is introduced and, if suitably
non-degenerate and satisfying a positivity condition, it is shown that
there is a canonically defined almost complex structure $J$ for which
the distribution is an almost complex contact distribution.
Furthermore, the obstruction to the integrability of $J$ is
identified. Our situation is somewhat different in that the primary
structure on a smooth manifold,
$P$, is a complex one-form $\tau$  defined up to complex
rescalings (or more abstractly, a complex line bundle $L^*\subset
\mathbbm{C} T^*P:=\IC\otimes T^*P$).
This is more information in the sense that $D$ is
defined directly as the kernel of $\tau$, but $\tau$ is only defined
by $D$ up to $\tau\mapsto a\tau + b\bar\tau$, where $a,b$ are
complex valued functions on $P$ . Given $D$, there is
a unique choice of $\tau$ that is compatible with the Cap-Eastwood
almost complex structure but a priori, one does not know if that is the
$\tau$ that has been chosen.  Our analogue of the
Cap-Eastwood theorem works in higher dimensions also and we state it
in greater generality than we need.

\begin{thm}\label{contactthm}
   Suppose that on a (smooth) manifold $P$ of dimension
   $4n+2$ we are given a complex line subbundle $L^*\subset
   \mathbbm{C} T^*P$, represented by a complex 
   one-form $\tau$ defined up to complex rescalings.
   Suppose further that
$$
   \tau\wedge(\d\tau)^{n+1}\ =\ 0 \qquad\mbox{and}\qquad
   \tau\wedge(\d\tau)^n\wedge \bar \tau \wedge (\d\bar\tau)^n\ \neq\ 0,
$$
   then there is a unique integrable almost complex structure
   for which $\tau$ is proportional to a non-degenerate holomorphic
   contact structure. Here, $(\d\tau)^n:=\d\tau\wedge\cdots\wedge\d\tau$
   ($n$-times).
\end{thm}

\proof We claim that, with the assumptions above, the
$(2n+1)$-form $\tau\wedge(\d\tau)^n$ is simple, i.e., that the space
of vectors $X\in\Gamma(P,\mathbbm{C}TP)$ such that $ X\hook
(\tau\wedge\d\tau)=0$ is $(2n+1)$-dimensional.  This follows because
the kernel of $\tau$ is $(4n+1)$-dimensional, whereas $\d\tau$ defines
a skew form on this kernel and so must have even rank.  However, its
rank is less than $2n+2$ by $\tau\wedge(\d\tau)^{n+1}=0$ but greater
than or equal to $2n$ because $\tau \wedge (\d\tau )^n\neq 0$.  Hence,
the kernel of $\tau \wedge (\d\tau)^n $ is $(2n+1)$-dimensional and we
will take this kernel to be the space of anti-holomorphic tangent
vectors spanning $T^{(0,1)}P$.  The condition that $T^{(0,1)}P$ should
contain no real vectors follows from the second assumption of the
theorem.

We have that
$X\hook(\tau\wedge(\d\tau)^n)=0 \Leftrightarrow
X\hook(\tau\wedge\d\tau)=0$ and we will use this latter
characterisation of $T^{(0,1)} P$ in the following.

We now consider the integrability of the distribution.  Let $X$ and $Y$
satisfy 
\begin{equation}
X\hook (\tau\wedge\d\tau)\ =\ 0\ =\ Y\hook (\tau\wedge\d\tau).
\end{equation}
Then clearly $X\hook\tau=0=Y\hook\tau$ and
\begin{equation}
\tau\wedge (X\hook \d\tau)\ =\ 0
\end{equation}
so that $X\hook\d\tau\propto \tau$ and $\mathcal{L}_X\tau\propto\tau$, and
similarly for $Y$. Here, $\mathcal{L}_X$ denotes the Lie derivative along $X$. Thus,
\begin{equation}
[X,Y]\hook\tau\ =\ X(Y\hook\tau)-Y(X\hook\tau)-X\hook(Y\hook\d\tau)\ =\ 0
\end{equation}
since $X\hook\tau=0=Y\hook\tau$ by assumption and so $X\hook(Y\hook\d\tau)=0$
from above. Furthermore,
\begin{eqnarray}
[X,Y]\hook (\tau\wedge\d\tau)\!\!&=&\!\! -\tau \wedge ([X,Y]\hook \d\tau)\notag\\
    \!\!&=&\!\!-\tau
\wedge ([X,Y]\hook \d\tau + \d ([X,Y]\hook\tau)\notag\\ \!\!&=&\!\! -\tau\wedge
(\mathcal{L}_{[X,Y]}\tau)\ =\ -\tau\wedge(\mathcal{L}_X \mathcal{L}_Y\tau-
   \mathcal{L}_Y\mathcal{L}_X\tau)\ =\ 0
\end{eqnarray}
since $\mathcal{L}_X\tau=X\hook\d\tau\propto\tau$, so
$\mathcal{L}_X\mathcal{L}_Y\tau\propto\tau$.

Thus, the almost complex structure is integrable.
\hfill $\Box$

\bigskip

In the twistor context, we will take $P$ to be a six-dimensional
manifold with topology $U\times S^2$ with $U\subset\mathbbm{R}^4$ and,
as before, we shall denote it by $PT$.
With this theorem, then, our data is simply a complex line subbundle
$L^*\subset \mathbbm{C} T^*PT$ represented by a differential one-form
$\tau$ with values in $L$ subject to the open condition $\tau \wedge
\d\tau\wedge \bar \tau \wedge\d\bar\tau\neq 0$.  We will
also require  that the line bundle $L$ has Chern class $2$.
The field equation is $\tau\wedge (\d\tau)^2=0$.  The $\mathcal{N}=0$
action above is simply
\begin{equation}
S[b,\tau]\ =\ \int b\wedge \tau\wedge(\d\tau)^2
\end{equation}
where $b\in\Omega^1PT\otimes (L^*)^{3}$ is a Lagrange multiplier.
Clearly, the field equation obtained by varying $b$ is $\tau\wedge
(\d\tau)^2=0$, as desired.
The action is clearly diffeomorphism
invariant, and enjoys a gauge invariance given by $\tau\mapsto\chi\tau$
and $b\mapsto\chi^{-3}b$, where $\chi$ is a non-vanishing
complex-valued function on $PT$.  This gauge freedom corresponds to
the fact that $\tau$ takes values in a line bundle $L$ which we shall
also denote by $\cO(2)$ since it becomes that on-shell, and hence $b$
is a differential one-form with values in $\cO(-6)$.

The action is also invariant under $b\mapsto b+\gamma$ where
$\gamma\wedge\tau\wedge(\d\tau)^2=0$, and the space of such $\gamma$
is two-dimensional when the field equations are not satisfied, but
three-dimensional when they are.  (When they are satisfied, this
freedom can be used to ensure that $b$ is a $(0,1)$-form.)  There is
also a gauge freedom in $b$ obtained as follows.  We can define a
partial connection $\bar\p$ on $\cO(n)$ by defining for $\chi$, now
assumed to be a section of $\cO(-6)$, $\bar\p\chi$ to be the differential
one-form modulo the kernel of $\bar\p \chi\mapsto \bar\p \chi\wedge
\tau\wedge(\d\tau)^2$ defined by $\bar\p
\chi\wedge\tau\wedge(\d\tau)^2:=\d (\chi\tau\wedge(\d\tau)^2)$.  It is
clear from this definition that the integrand of the action evaluated
on such a $b=\bar\p \chi$ is a boundary integral and so this represents a gauge
freedom.  On-shell, the above definition becomes trivial, and
$\bar\p\chi$ needs to be defined a little differently by $\bar\p
\chi^{2/3} \wedge (\tau\wedge\d\tau):= \d
(\chi^{2/3}\tau\wedge\d\tau)$ and in this case it leads to an honest
$\bar\p$-operator on the line bundles $\cO(n)$.

The field equation for $b$ is
\begin{equation}
\d b\wedge \tau \wedge \d\tau - {\textstyle\frac 32} b\wedge(\d\tau)^2\ =\ 0
\end{equation}
and when the field equation for $\tau$ is satisfied, this is the
$\bar\p$-closure condition for sections of
$\Omega^{(0,1)}PT\otimes\cO(-6)$.  Taking into account the gauge freedom
$b\mapsto b+\bar\p \chi$ with $\chi$ a section of $\cO(-6)$, $b$ will
correspond to an element of $H^1(PT, \cO(-6))$.

Thus, solutions to the field equations correspond to a complex
three-dimensional
manifold $PT$ with holomorphic contact structure $\tau$, and the
condition on the Chern class of $L$ implies that it satisfies the
topological assumption of Ward's theorem, so that, if it contains a
holomorphic rational curve of degree one in the $S^2$-factor, then it
corresponds to a space-time $M$ with self-dual Einstein metric. The
field $b\in H^1(PT,\cO(-6))$ then corresponds via the Penrose
transform to a right-handed linearised gravitational field propagating
on that self-dual background.  Thus, we have the self-dual sector of
non-supersymmetric Einstein gravity.

\subsection{The supersymmetric case}

In the supersymmetric situation, we will assume that $\cP\cT$ is a
smooth supermanifold with six real bosonic dimensions and $\CN$
complex fermionic dimensions.  Without loss of generality, we can
always assume that the supermanifold is split in the smooth category
(Batchelor (1979)), and that locally the odd coordinates are $\theta^i$,
$i=1,\ldots,\CN$, and that we will only ever have holomorphic
dependence on $\theta^i$, their complex conjugates will not enter the
formalism, so, in particular, the transition functions for the
supermanifold will be holomorphic in $\theta^i$.\footnote{In a Dolbeault
  context, this assumption is, in effect a gauge choice.} We can still
encode the structure of a supersymmetric non-linear graviton into a
complex contact form $\tau$ as follows.  We will assume that $\tau$ is
a complex differential one-form on the supermanifold $\cP\cT$, again
with only holomorphic dependence on the $\theta^i$, i.e., $\tau=
\d x^a\tau_a + \d\theta^i\tau_i$, where the $x^a$s are the real bosonic
coordinates on $\cP\cT$, $a=1,\ldots, 6$, and $\tau_a$ and $\tau_i$
are holomorphic in $\theta^i$ with $\tau_i$ odd and $\tau_a$ even
functions on $\cP\cT$.  On the body of the supermanifold, $\theta^i=0$,
we can assume that we have the equations $\tau\wedge(\d\tau)^2=0$ as
before, but these will not hold when $\theta^i\neq 0$, even for standard
flat supertwistor space as, in general, $(\d\theta)^n\neq 0$ $\forall\
n$ for an odd variable $\theta$.  Thus, we cannot express the conditions
we need quite so simply in the supersymmetric case.

Nevertheless, much of Thm.~\ref{contactthm} works in the
supersymmetric case also.  We will require firstly, as a genericity
assumption, that the complexified kernel $\mathbbm{C}\cD$ of $\tau$
has dimension $5|2\CN$ (here we are taking $\p/\p\theta^i$ and
$\p/\p\bar\theta^{\bar i}$ to be independent). Secondly, we require that on this
complexified kernel of $\tau$, the two form $\d\tau$ has rank $2|\CN$
so that the kernel of $\tau\wedge\d\tau$ is $3|\CN$-dimensional
and further, that $\ker(\tau\wedge\d\tau)$ has no real vectors, i.e.
\begin{equation}
\ker(\tau\wedge\d\tau)\cap\overline{\ker(\tau\wedge\d\tau)}\ =\ \{0\}.
\end{equation}
The fact that we have required that $\tau$ depends only on $\theta^i$ and not
$\bar\theta^{\bar i}$ means that $\d\tau$ annihilates $\p/\p\bar\theta^{\bar i}$, for
$i=1,\ldots,\CN$ and so the rank of $\d\tau$ is at most $5|\CN$ in any
case.  With these assumptions, the proof of Thm.~\ref{contactthm}
follows without modification to show that $\ker(\tau\wedge\d\tau)$
is integrable and that $\tau$ is a holomorphic complex contact
structure so that
\begin{equation}
T^{(0,1)}\cP\cT\ :=\ \ker(\tau\wedge\d\tau).
\end{equation}

The main field equation is therefore the condition that
$\tau\wedge\d\tau$ annihilates a complex distribution of dimension
$3|\CN$.  In the supersymmetric context, we do not yet have an
equation on $\tau$ analogous to the bosonic equation
$\tau\wedge(\d\tau)^{n+1}=0$ for higher dimensional complex contact
structures nor an action that produces this condition as its
Euler-Lagrange equation.  As a consequence, we have so far been unable
to find a covariant supersymmetric action functional.

\section{Conclusions}\label{conclusions}

Given that these actions are `Chern-Simons-like' one is led to ask the
extent to which they can be interpreted coherently as holomorphic
Chern-Simons theories.  Clearly, in some sense, the gauge group should
be taken to be the diffeomorphisms of the supertwistor space that
preserve the holomorphic Poisson structure.  This is most easily made
sense of in a complexified context so that the holomorphic twistor
variables are freed up and become independent from the conjugate
twistor variables.  Then the theory becomes a complexified
Chern-Simons theory with gauge group the holomorphic contact
transformations of the holomorphic supertwistor space, a region in
$\CP^{3|8}$, on the conjugate supertwistor space (which is just
$\CP^3$ as we have no anti-holomorphic fermionic coordinates).  A
similar connection between the self-dual vacuum equations and a gauge
theory with a diffeomorphism group gauge group was given on space-time
in Mason \& Newman (1989) (here the gauge theory was the self-dual
Yang-Mills equations); see also Wolf (2007) for a supersymmetric
extension thereof.

The fact that Thm.~\ref{contactthm} works in $4n+2$ dimensions is
suggestive of applications of this framework to the twistor theory for
quaternionic K\"ahler manifolds with non-zero scalar curvature in $4n$
dimensions.  It is straightforward to write down a Lagrange multiplier
action $\int b\wedge\tau\wedge(\d\tau)^{n+1}$ analogous to our $\CN=0$
action, but with $b$ a $(2n-1)$-form, although in this context the
interpretation of $b$ is less clear.

An attractive feature is that we have a fully supersymmetrically
invariant and Lorentz invariant off-shell formulation of the theory.
However, we have so far been unable
to find an action functional of $\CN=8$ self-dual supergravity that
does not depend on a given integrable background.  Such an action
functional would, however, be desirable as one would hope for an
explicitly diffeomorphism invariant action principle for $\CN=8$
self-dual supergravity.  In particular, if one wishes to be able to
extend the ideas to the full theory along the lines of Mason (2005)
for conformal supergravity,\footnote{See also Abou-Zeid \& Hull (2006)
  for a space-time action for expanding about the self-dual sector in
  the case of Einstein gravity.} then it would seem awkward to have to
identify a Minkowski background.

A task for the future is to start with the superfield expansions (in the
non-linear setting) of $\tau$ and $h$ and reproduce the covariant form
of the field equations and of the action functional of $\CN=8$
self-dual supergravity in four dimensions as given in Siegel
(1992).\footnote{Similar expansions for certain supersymmetric gauge
  theories were performed in Popov \& Wolf (2004), Popov \& S\"amann
  (2005), S\"amann (2005), Popov, S\"amann \& Wolf (2005) and
  Lechtenfeld \& S\"amann (2006).}  In the zero cosmological constant
case, our twistor action and field equations must correspond via the
Penrose transform to Siegel's results.  
\\

\noindent
{\bf Acknowledgements.}
We would like to thank Alexander Popov for a number of important
contributions to this work and Mohab Abou-Zeid, Rutger Boels, Daniel
Fox, Chris Hull, Riccardo Ricci, Christian S\"amann and David Skinner
for useful discussions. We would also like to thank the referee for
useful suggestions.  The first author is partially supported by the EU
through the FP6 Marie Curie RTN {\it ENIGMA} (contract number
MRTN--CT--2004--5652) and through the ESF MISGAM network. The second
author was supported in part by the EU under the MRTN contract
MRTN--CT--2004--005104 and by STFC under the rolling grant
PP/D0744X/1.

\appendix
\section{Prepotential formulation}\label{sec:app-pf}

The subject of this appendix is the comparison of Karnas' \& Ketov's
(1998) approach with ours. Their formulation is based on an
anti-holomorphic involution which picks a real slice in complexified
space-time being of split signature. Pretty much the same holds true,
however, for Euclidean signature and it is this latter case we are
interested in here. As already indicated, this works only for an even
number of supersymmetries.  In the following, we shall use 
conventions from Wolf (2006).

\subsection{Real structures on $\IP\IT'_{[\CN]}$ and $\IM_{[\CN]}$}

Let us first consider the supertwistor space
$\IP\IT'_{[\CN]}=\CP^{3|\CN}\setminus\CP^{1|\CN}$ with (homogeneous)
coordinates $(\omega^A,\pi_{\dot\alpha})$ for flat super space-time
  $\IM_{[\CN]}\cong\IC^{4|2\CN}$.
An Euclidean signature real slice follows from the anti-holomorphic involution
without fixed points $\rho\,:\,\IP\IT'_{[\CN]}\to\IP\IT'_{[\CN]}$ given by
\begin{equation}
(\hat \omega^A, \hat\pi_{\dot\alpha})\ :=\
\rho(\omega^A,\pi_{\dot\alpha})\ :=\ (\bar\omega^B{C_B}^A,
  {C_{\dot\alpha}}^{\dot\beta}\bar\pi_{\dot\beta}),
\end{equation}
where bar denotes complex conjugation and
$({C_A}^B)=\mbox{diag}(({C_\alpha}^\beta),({C_i}^j))$, with
\begin{equation}
  ({C_\alpha}^\beta)\ =\ \epsilon,\quad
  ({C_i}^j)\ =\ \mbox{diag}(\!\!\underbrace{\epsilon,
   \ldots,\epsilon}_{\ \frac{\CN}{2}{\rm-times}}\!\!),\quad
   ({C_{\dot\alpha}}^{\dot\beta})\ =\ -\epsilon,\quad
   \epsilon\ :=\ \begin{pmatrix}
                  0 & 1\\ -1 & 0
                 \end{pmatrix}\!.
\end{equation}
We can extend $\rho$ to a map from a holomorphic function $f$ on
$\IP\IT'_{[\CN]}$ another holomorphic function by
\begin{equation} 
\rho(f(\cdots))\ :=\ \overline{f(\rho(\cdots))}.
\end{equation}
By virtue of the incidence relation, $\omega^A=x^{A\dot\alpha}\pi_{\dot\alpha}$,
we obtain an induced involution on $\IM_{[\CN]}$ explicitly
given by
\begin{equation}
 \rho(x^{A\dot\alpha})\ =\ -\bar x^{B\dot\beta}{C_B}^A{C_{\dot\beta}}^{\dot\alpha}.
\end{equation}
We shall use the same notation $\rho$ for the
anti-holomorphic involution induced on the
different (super)manifolds in the twistor correspondence.
The fixed point set of this involution, that is,
$\rho(x)=x$ for $x\in\IM_{[\CN]}$, defines Euclidean right-chiral
superspace $\IM^\rho_{[\CN]}\cong\IR^{4|2\CN}$ inside $\IM_{[\CN]}$.

Following Atiyah, Hitchin \& Singer (1978), the supertwistor space
$\IP\IT'_{[\CN]}$ can be
identified with
\begin{equation}
 \cO(1)^{\oplus 2|\CN}\ \to\ \CP^1
\end{equation} 
and so it can be covered
by two (acyclic) coordinate patches $\cU_\pm$ and coordinatised
by $(\omega^A_\pm,\pi_\pm)$, where $\omega^A_\pm$
are local fibre coordinates with $\omega^A_+:=\omega^A/\pi_{\dot0}$,
$\omega^A_-:=\omega^A/\pi_{\dot1}$ and $\pi_+:=\pi_{\dot1}/\pi_{\dot0}$,
$\pi_-:=\pi_{\dot0}/\pi_{\dot1}$ are the standard
local holomorphic coordinates on $\CP^1$, with $\pi_+=\pi_-^{-1}$ on
$\cU_+\cap\cU_+\subset\IP\IT'_{[\CN]}$.
On the other hand, since $\IP\IT'_{[\CN]}$ is diffeomorphic to
$\IM^\rho_{[\CN]}\times S^2\cong\IR^{4|2\CN}\times S^2$, one may
equivalently coordinatise it by using
$(x^{A\dot\alpha},\lambda_\pm)$, where $\lambda_\pm$
are the standard local holomorphic coordinates on $S^2\cong\CP^1$.
Note that $(\omega^A_\pm,\pi_\pm)=(x^{A\dot\alpha}
\lambda^\pm_{\dot\alpha},\lambda_\pm)$, where
\begin{equation}\label{eq:deflam}
(\lambda^{\dot\alpha}_+)\ :=\
     \begin{pmatrix}\lambda_+\\-1\end{pmatrix}\qquad\mbox{and}\qquad
    (\lambda^{\dot\alpha}_-)\ :=\
     \begin{pmatrix}1\\-\lambda_-\end{pmatrix}.
\end{equation}
The explicit inverse transformation laws are
simply
\begin{equation}
x^{A\dot\alpha}\ =\  \frac{\omega^A_\pm \hat\pi_\pm^{\dot\alpha}-
  \hat\omega^A_\pm \pi_\pm^{\dot\alpha}}{\hat\pi_\pm^{\dot\beta}\pi_{\pm\dot\beta}},
\end{equation}
where $\pi^{\dot\alpha}_\pm$ are similarly defined as in \eqref{eq:deflam}.
Altogether, we have obtained a non-holomorphic
fibration
\begin{equation}
\pi\,:\,\IP\IT'_{[\CN]}\ \to\ \IM^\rho_{[\CN]}.
\end{equation}

Introduce
\begin{equation}
(\hat\lambda^{\dot\alpha}_+)\ :=\
      \begin{pmatrix} 1\\ \bar\lambda_+\end{pmatrix}\!,\quad
    (\hat\lambda^{\dot\alpha}_-)\ :=\
      \begin{pmatrix}\bar\lambda_-\\1\end{pmatrix}\!,
    \quad\gamma_\pm^{-1}\ :=\
    \hat\lambda^{\dot\alpha}_\pm
     \lambda^\pm_{\dot\alpha}\ =\ 1+\lambda_\pm\bar\lambda_\pm,
\end{equation}
like for $\hat\pi^{\dot\alpha}_\pm=\rho(\pi^{\dot\alpha}_\pm)$.
Then, due to the above diffeomorphism, we have the following
transformation laws between the coordinate vector fields:
\begin{subequations}
\begin{eqnarray}
  \frac{\p}{\p\omega^A_\pm}\!\!&=&\!\!\gamma_\pm\hat
   \lambda^{\dot\alpha}_\pm\frac{\p}{\p x^{A\dot\alpha}},\\
  \frac{\p}{\p\pi_+}\!\!&=&\!\!\frac{\p}{\p\lambda_+}-\gamma_+
   x^{A\dot1}\hat\lambda^{\dot\alpha}_+
  \frac{\p}{\p x^{A\dot\alpha}},\\
\frac{\p}{\p\pi_-}\!\!&=&\!\!\frac{\p}{\p\lambda_-}-\gamma_-
  x^{A\dot0}\hat\lambda^{\dot\alpha}_-
  \frac{\p}{\p x^{A\dot\alpha}}
\end{eqnarray}
for the holomorphic tangent vector fields and
\begin{eqnarray}
  \frac{\p}{\p\bar\omega^{\bar A}_\pm}\!\!&=&\!\! -
  \gamma_\pm{C_A}^B\lambda^{\dot\alpha}_\pm\frac{\p}{\p x^{B\dot\alpha}},\\
  \frac{\p}{\p\bar\pi_+}\!\!&=&\!\! \frac{\p}{\p\bar\lambda_+}-
   \gamma_+ x^{A\dot0}\lambda^{\dot\alpha}_+
  \frac{\p}{\p x^{A\dot\alpha}},\\
\frac{\p}{\p\bar\pi_-}\!\!&=&\!\!\frac{\p}{\p\bar\lambda_-}+\gamma_-
  x^{A\dot1}\lambda^{\dot\alpha}_-
  \frac{\p}{\p x^{A\dot\alpha}}
\end{eqnarray}
\end{subequations}
for the anti-holomorphic ones.

\subsection{Comparison of the two approaches}

In what follows, we shall restrict our discussion to the $\cU_+$-patch
only and for notational simplicity suppress the patch index. Of course,
a similar discussion carries over to the $\cU_-$-patch.

To begin with, let us write down the field equations
\eqref{eq:fieldeq} more explicitly.
If we let the deformation be $h=\d\bar\omega^{\bar\alpha}
h_{\bar\alpha}+\d\bar\pi h_{\bar\pi}$, they read as
\begin{subequations}
\begin{eqnarray}
  \frac{\p}{\p\bar\omega^{\bar\alpha}} h_{\bar\beta}-
  \frac{\p}{\p\bar\omega^{\bar\beta}} h_{\bar\alpha}+
  [h_{\bar\alpha},h_{\bar\beta}\}\!\!&=&\!\!0,\\
  \frac{\p}{\p\bar\pi}h_{\bar\alpha}-
  \frac{\p}{\p\bar\omega^{\bar\alpha}}h_{\bar\pi}+
   [h_{\bar\pi},h_{\bar\alpha}\}\!\!&=&\!\!0.
\end{eqnarray}
\end{subequations}
Using the incidence
relation $\omega^A=x^{A\dot\alpha}\pi_{\dot\alpha}$ and
the involutions introduced in the preceding subsection, $h$ can
also be expressed in the coordinates $(x^{A\dot\alpha},\lambda)$
as
\begin{equation}
h\ =\ -\gamma\hat\lambda_{\dot\beta}\d x^{\alpha\dot\beta}\,
  \Phi_{\alpha}+\d\bar\lambda\,\Phi_{\bar\lambda},
\end{equation}
where $\Phi_{\alpha}:=-\gamma^{-1}{C_\alpha}^\beta h_{\bar\beta}$
and $\Phi_{\bar\lambda}:=
h_{\bar\pi}+\gamma x^{\alpha\dot0}\Phi_{\alpha}$.

In order to compare our approach with those by Karnas \& Ketov (1998),
we notice that their formulation deals with the `vacuum case',
i.e.~with the case of vanishing cosmological constant. 
Upon also recalling point (iii) of Thm.~\ref{wardthm}, we must therefore ensure that
the fibration of the supertwistor space is preserved,
and so (i) $h$ is of the form
\begin{equation}\label{eq:hzero}
h\ =\ -\gamma\hat\lambda_{\dot\beta}\d x^{\alpha\dot\beta}
  \,\Phi_{\alpha},
\end{equation}
i.e. $\Phi_{\bar\lambda}=0$ $\Leftrightarrow$
$h_{\bar\pi}=-\gamma x^{\alpha\dot0}\Phi_{\alpha}$
and (ii) the relative symplectic structure needs to be
preserved which amounts to requiring a degeneracy of the
Poisson structure $\omega=( I^{IJ})$ introduced in \S\ref{nonlinext}
according to $\omega=( I^{AB})$. Notice further that
$\Phi_\alpha$ must be of weight $3$ in order for $h$ to be of weight
$2$.

Some algebra then reveals that in the `vacuum case' the above equations for
$h_{\bar\alpha}$ and $h_{\bar\pi}$ translate into the following
set:
\begin{subequations}
\begin{eqnarray}
  \epsilon^{\alpha\beta}\dbar_\alpha\Phi_\beta+\textstyle{\frac{1}{2}}
   \epsilon^{\alpha\beta}[\Phi_\alpha,\Phi_\beta\}\!\!&=&\!\!0,\\
  \p_{\bar\lambda}\Phi_\alpha+\gamma^{-2}\epsilon^{\beta\gamma}
  (\partial_\beta\Phi_\alpha)\Phi_\gamma\!\!&=&\!\!0,
\end{eqnarray}
\end{subequations}
where $\dbar_A:=\lambda^{\dot\alpha}\p/\p x^{A\dot\alpha}$
and $\p_A:=\gamma\hat\lambda^{\dot\alpha}\p/\p x^{A\dot\alpha}$.

Before going any further, let us say a few words about gauge symmetries.
The original equations for $h$ transformed covariantly under
gauge transformations of the form $h\mapsto h+\delta h$,
with $\delta h=\dbar_0 \chi+[h,\chi\}$ for some function $\chi$
of weight 2. However, the above equations will no longer
transform covariantly under generic gauge transformations,
since we have incorporated the
constraint $\Phi_{\bar\lambda}=0$. Nevertheless,
some residual gauge symmetry remains, and which is determined as follows.
In order to preserve the constraint $\Phi_{\bar\lambda}=0$,
we must have
$\delta h_{\bar\pi}=
-\gamma x^{\alpha\dot0}\delta\Phi_{\alpha}$, where
$\delta\Phi_\alpha=-\gamma^{-1}{C_\alpha}^\beta\delta
h_{\bar\beta}$, i.e. transformations of $h_{\bar\pi}$
are determined by those of $h_{\bar\alpha}$. It is not
difficult to verify that the remaining gauge symmetry
is given by the following transformation laws
\begin{equation}
 \delta\Phi_\alpha\ =\ -(\dbar_\alpha\chi+[\Phi_\alpha,\chi\}),
\qquad\mbox{with}\qquad
\p_{\bar\lambda}\chi+\gamma^{-2}\epsilon^{\beta\gamma}
  (\partial_\beta\chi)\Phi_\gamma\ =\ 0. 
\end{equation}

In particular, the last of these equations shows that the 2nd
equation for $\Phi_\alpha$ from above does not
constrain $\Phi_\alpha$ any further, so that the only
remaining field equation we are left with is
\begin{equation}
 \epsilon^{\alpha\beta}\dbar_\alpha\Phi_\beta+\textstyle{\frac{1}{2}}
   \epsilon^{\alpha\beta}[\Phi_\alpha,\Phi_\beta\}\ =\ 0.
\end{equation}
Since in particular $\Phi_\alpha=\p_\alpha\Phi$ (see also
Woodhouse (1985)), where
$\Phi$ is some function of weight 4 (recall that $\Phi_\alpha$ is
of weight 3) and $\omega=( I^{AB})$, we end up with
\begin{equation}\label{eq:fieldeqKK}
\Box\Phi+\textstyle{\frac{1}{2}}
   \epsilon^{\alpha\beta}(-)^{p_A}\p_A\p_\alpha\Phi\, I^{AB}\,\p_B
\p_\beta\Phi\ =\ 0\qquad\mbox{and}\qquad \Box\ :=\
\epsilon^{\alpha\beta}\dbar_\alpha\p_\beta,
\end{equation}
which is Karnas' \& Ketov's (1998) result. 

As before, in the case of maximal supersymmetry, $\CN=8$, the
field equations \eqref{eq:fieldeqKK} can be derived from an action
principle,
\begin {subequations}
\begin{equation}
 S[\Phi]\ =\ \int \d\,{\rm vol}
\left\{\Phi\Box\Phi+\textstyle{\frac{1}{3!}}
   \epsilon^{\alpha\beta}(-)^{p_A}\Phi\p_A\p_\alpha\Phi\, I^{AB}\,\p_B
\p_\beta\Phi\right\}\!,
\end{equation}
where the measure $\d\,{\rm vol}$ is given by
\begin{equation}
 \d\, {\rm vol}\ =\ \d^4x\,\gamma^2\d\lambda\d\bar\lambda\,\d\theta^1\cdots\d\theta^8.
\end{equation}
\end{subequations}

It remains to give the superfield expansion of $\Phi$. For brevity,
let us only discuss the $\CN=8$ case. We find
\begin{eqnarray}
 \Phi\!\!&=&\!\! g+\theta^i \psi_i + \theta^{i_1i_2}A_{[i_1i_2]}+ \theta^{i_1i_2i_3}\chi_{i_1i_2i_3}+
    \theta^{i_1i_2i_3i_4}\phi_{i_1i_2i_3i_4}\ +\notag\\
     &&\kern1cm+\ \theta_{i_1i_2i_3}\tilde\chi^{i_1i_2i_3}+\theta_{i_1i_2}\tilde A^{i_1i_2}+
     \theta_i\tilde\psi^i+\theta\tilde g,          
\end{eqnarray}
where 
\begin{subequations}
\begin{eqnarray}
 \theta^{i_1\cdots i_r}\!\! &:=&\!\! 
  {\textstyle\frac{1}{r!}}\theta^{i_1}\cdots\theta^{i_r},\quad\mbox{for}\quad r=1,\ldots,4,\\
 \theta_{i_1\cdots i_{8-r}}\!\! &:=&\!\! {\textstyle\frac{1}{r!}}\epsilon_{i_1\cdots i_{8-r}i_{9-r}
 \cdots i_{8}}\theta^{i_{9-r}}\cdots\theta^{i_8},\quad\mbox{for}\quad r=5,\ldots,8.
\end{eqnarray}
\end{subequations}
Here, $\epsilon_{i_1\cdots i_8}=\epsilon_{[i_1\cdots i_8]}$ and
$\epsilon_{1\cdots 8}=1$.
Keeping in mind \eqref{eq:hzero}, we find
the following space-time fields:
\begin{center}
\let\PBS=\PreserveBackslash
\setlength{\extrarowheight}{4pt}
 \begin{tabular}{| >{\PBS\raggedright\hspace{0pt}}p{2cm} 
                 | >{}p{.5cm}| >{}p{.5cm}|>{}p{.5cm}|>{}p{.5cm}|>{}p{.5cm}|
               >{}p{.5cm}|>{}p{.5cm}|>{}p{.5cm}|>{}p{.5cm}|}  
 \hline
      Field              
        & $g$ & $\psi$ & $A$ & $\chi$ & $\phi$ & $\tilde\chi$ & $\tilde A$ & $\tilde\phi$ & $\tilde g$
            \\
\hline
      Helicity & 2 & ${\textstyle\frac32}$ & 1 & ${\textstyle\frac12}$ & 0 & 
             $-{\textstyle\frac12}$ & $-1$ & $-{\textstyle\frac32}$ & $-2$ \\
\hline
      Multiplicity & 1 & 8 & 28 & 56 & 70 & 56 & 28 & 8 & 1\\
\hline 
 \end{tabular}

\vspace*{10pt}
{\small {\bf Table 1:} Space-time fields and their helicities and multiplicities.}
\end{center}

\section{Holomorphic volume forms and non-projective twistor
  space}\label{nonproj} 
It is often convenient to work on non-projective twistor space $\IT$ as
many of the geometric structures can be formulated globally there and
sections of the line bundles $\cO(n)$ become ordinary functions of
weight $n$ under the action of the Euler vector field
$\Upsilon=Z^I\p/\p Z^I$.  In the curved case, as in the proof of
theorem \ref{wardthm}, the non-projective space can be defined as the
quotient of the non-projective co-spin bundle $\cS^*$ by $\cD_\cF$.
We can also define it intrinsically as follows.

In the bosonic case, given a contact structure defined by a
one-form
$\tau$ with values in a line bundle $L$, we can see that
$\tau\wedge\d\tau$ defines a (non-vanishing)
section of $\Omega^{(3,0)}PT\otimes
L^{2}$.  Thus, we must have $L^{-2}\cong\Omega^{(3,0)}PT$.  In the flat case,
non-projective twistor space $\IT_{[0]}\cong\IC^{4}$ is the total space of the
(tautological) line bundle $\cO(-1)$ over the projective twistor
space $\IP\IT'_{[0]}$, and $\Omega^{(3,0)}\IP\IT'_{[0]}\cong\cO(-4)$.
In the general (non-supersymmetric) case, we
can define the non-projective twistor space $T$ to be the total space
of the line bundle $\cO(-1)$ now defined to be the 4th root of
$\Omega^{(3,0)}PT$.  If so, we see that $L\cong\cO(2)$.  The
non-projective space has an Euler vector field $\Upsilon$ that
generates the $\IC^*$ action on the fibres of $\cO(-1)$. The weights
of functions and forms pulled back from $PT$ are translated into the
weights along $\Upsilon$ on the non-projective space.  In this
context, $\tau$ defines a 1-form of weight 2 on the non-projective
space, and the non-degeneracy of the contact structure translates into
the condition that the two-form $\d \tau$ is non-degenerate as a
two-form on $T$ and being closed defines a holomorphic symplectic
structure.  Its inverse $\Pi$ therefore defines a non-degenerate holomorphic
Poisson structure on $T$ of weight $-2$.  This descends to give a
Poisson structure on $PT$with values in $\cO(-2)$.     

We can extend this reasoning to the supersymmetric case as follows.
We again consider a holomorphic differential one-form $\tau$ with
values in a complex line bundle $\cL$.  It defines as its kernel
the contact distribution $\cD$, which now is of rank $2|\CN$,
leading to a short exact sequence as follows:
\begin{equation}
   0\ \longrightarrow\ \cD\ \longrightarrow\
    T^{(1,0)}\cP\cT\ \longrightarrow\ \cL\
   \longrightarrow\ 0\, .
\end{equation}
Since we
assume that $\tau$ defines a non-degenerate holomorphic contact structure,
$\d\tau$ provides a non-degenerate skew form on $\cD$. Taking its
Berezinian, we get an element
\begin{equation}
{\rm Ber}(\d\tau|_\cD)\
\in\ \cL^{2-\CN}\otimes ({\rm Ber}\,\cD)^{-2}.
\end{equation}
(This follows from the fact that in the definition of the Berezinian,
the odd-odd part of the matrix is inverted before its determinant is
taken leading to inverse weights associated to the odd directions
relative to their bosonic counterparts.)
When $\cL$ has a square root, we can take its square root to get an
isomorphism
\begin{equation}
\sqrt{({\rm Ber}(\d\tau|_\cD))}\,:\, {\rm
    Ber}\,\cD\ \to\ \cL^{1-\CN/2}\, .
\end{equation}
The above exact sequence then gives an identification
\begin{equation}
{\rm Ber}\,T^{(1,0)}\cP\cT\ \cong\
{\rm Ber}\,\cD\otimes \cL\ \cong\ \cL^{2-\CN/2}
\end{equation}
and so finally we obtain the isomorphism
\begin{equation}\label{eq:ber}
{\rm Ber}(\cP\cT)\
   :=\ {\rm Ber}\,\Omega^{(1,0)}\cP\cT\ \cong\ \cL^{\CN/2-2}.
\end{equation}

We will take the body of the supertwistor space to have topology
$U\times S^2$ where $U$ is an open subset of $\mathbbm{R}^4$
(or more generally the total space of the projective co-spin bundle
of a real smooth spin four-manifold $M$).
The assumption on the normal bundle
of a rational curve in supertwistor space implies that the
holomorphic Berezinian
bundle ${\rm Ber}(\cP\cT)$ has Chern class $\CN-4$, and with the topological
assumptions we have made, this will have an $|\CN-4|$-th root and we
may introduce the (consistent) notation
$\cO(n):=({\rm Ber}(\cP\cT))^{n/(\CN-4)}$. Thus, $\cL\cong\cO(2)$ and
${\rm Ber}(\cP\cT)\cong\cO(\CN-4)$.

\section{Supersymmetric BF-type theory}\label{sec:app-superbf}

In this appendix we wish to present an alternative interpretion
of the holomorphic Chern-Simons-type theory \eqref{eq:csaction}.
We shall see that this theory can be viewed as a certain
supersymmetric holomorphic BF-type theory. In what follows, we 
will borrow ideas of Witten (1989).

To begin with, consider some $(0|2)$-dimensional space $\cT$ with odd
coordinates $\psi^1$ and $\psi^2$, which we collectively denote
by $\psi^\alpha$. On $\cP\cT\times\cT$, we
may introduce a $(0,1)$-form $H$ of weight 2 according to
\begin{equation}
 H\ =\ h+\psi^\alpha\chi_\alpha+\psi^1\psi^2 b.
\end{equation}
Here, $h$ and $b$ are even and $\chi_\alpha$ are odd
$(0,1)$-forms of weight 2 on $\cP\cT$.
As before, we assume that these fields
have no dependence on the $\bar \theta^{\bar i}$ coordinates.

In analogy to \eqref{eq:csaction}, we may consider the action functional
\begin{equation}
 S[b,h,\chi_\alpha]\ =\  \int\d\psi^1\d\psi^2\int\Omega_8^{(0)}\wedge 
               \left(H\wedge\bar\p_0 H + {\textstyle\frac 13}H\wedge[H,H\}\right)\!.
\end{equation}
A short calculation reveals that this action reduces after integration
over the $\psi^\alpha$ coordinates to
\begin{equation}\label{eq:susybf}
 S[h,b,\chi_\alpha]\ =\  \int\Omega_8^{(0)}\wedge 
               \left\{b\wedge F^{(0,2)} - {\textstyle\frac12}\epsilon^{\alpha\beta}
             \chi_\alpha\wedge (\bar\p_0\chi_\beta+[h,\chi_\beta\})\right\}\!.
\end{equation}
The equations of motion that follow from this action are
\begin{subequations}
\begin{eqnarray}
 F^{(0,2)} \!\!&=&\!\! 0,\\
 \bar\p_0 b+[h,b\} \!\! &=&\!\! {\textstyle\frac12}\epsilon^{\alpha\beta}
             [\chi_\alpha,\chi_\beta\},\\
\bar\p_0 \chi_\alpha+[h,\chi_\alpha\}\!\!&=&\!\!0.
\end{eqnarray}
\end{subequations}
The first equation is the field equation \eqref{eq:fieldeq}.
Note that for
$\chi_\alpha=0$ we get \eqref{eq:Bfaction}.

The supersymmetry transformations are straightforwardly worked out
as they follow from infinitesimal translations in the odd coordinates
$\psi^\alpha$. We find
\begin{equation}
   \delta_\alpha h\ =\ \chi_\alpha,\qquad 
               \delta_\alpha \chi_\beta\ =\ \epsilon_{\alpha\beta} b
             \qquad\mbox{and}\qquad
               \delta_\alpha b\ =\ 0,
\end{equation}
with $\{\delta_\alpha,\delta_\beta\}=0$. Therefore, the supersymmetric
holomorphic BF-type action \eqref{eq:susybf} can also be written as
\begin{equation}
  S[h,b,\chi_\alpha]\ =\ -{\textstyle\frac12}\delta_1\delta_2 S[h],
\end{equation}
where $S[h]$ is the action \eqref{eq:csaction}. 

%\newpage

\section*{References}

{\small

\smallskip
\noindent
    Abou-Zeid, M. and Hull, C.~M. (2006):
    {\it A chiral perturbation expansion for gravity},
    JHEP\ {\bf 0602}\ 057
    [arXiv:hep-th/0511189].
    %%CITATION = JHEPA,0602,057;%%

\smallskip
\noindent
    \rule{1cm}{.5pt}, \rule{1cm}{.5pt} and Mason, L.~J. (2008):
    {\it Einstein supergravity and new twistor string theories},
    Commun.\ Math.\ Phys.\  {\bf 282}\ 519
    [arXiv:hep-th/0606272].
    %%CITATION = HEP-TH 0606272;%%

\smallskip
\noindent
    Atiyah, M.~F., Hitchin, N.~J. \& Singer, I.~M. (1978):
    {\it Self-duality in four-dimensional Riemannian geometry}, 
    Proc.\ Roy.\ Soc.\ Lond.\ {\bf A 362}\ 425.

\smallskip
\noindent
Bailey, T.N.\ and Eastwood, M.G.\ (1991): {\it Complex  paraconformal
  manifolds--- their differential geometry and twistor theory},
Forum.\ Math.\ {\bf 3}\  61.

\smallskip
\noindent
    Batchelor, M. (1979):
    {\it The structure of supermanifolds},
    Trans.\ Amer.\ Math.\ Soc.\ {\bf 253}\ 329.

\smallskip
\noindent
    Bergshoeff, E. and Sezgin, E. (1992):
    {\it Self-dual supergravity theories in $(2+2)$-dimensions},
    Phys.\ Lett.\  B\ {\bf 292}\ 87
    [arXiv:hep-th/9206101].
    %%CITATION = HEP-TH 9206101;%%

\smallskip
\noindent
    Berkovits, N. (2004):
    {\it An alternative string theory in twistor space for $\CN=4$ super Yang-Mills},
    Phys.\ Rev.\ Lett.\ {\bf 93}\ 011601
    [arXiv:hep-th/0402045].
    %%CITATION = HEP-TH 0402045;%%

\smallskip
\noindent
    \rule{1cm}{.5pt} and Witten, E. (2004):
    {\it Conformal supergravity in twistor-string theory},
    JHEP\ {\bf 0408}\ 009
    [arXiv:hep-th/0406051].
    %%CITATION = HEP-TH 0406051;%%

\smallskip
\noindent
    Bern, Z., Dixon, L.~J. and Roiban, R. (2007):
    {\it Is $\CN=8$ supergravity ultraviolet finite?},
    Phys. Lett.\ B\ {\bf 644} 265
    [arXiv:hep-th/0611086].
    %%CITATION = HEP-TH 0611086;%%

\smallskip
\noindent
    Bershadsky, M., Cecotti, S., Ooguri H., and Vafa, C. (1994):
    {\it Kodaira-Spencer theory of gravity and exact results for quantum string amplitudes},
    Commun.\ Math.\ Phys.\ {\bf 165}\ 311
    [arXiv:hep-th/9309140].
    %%CITATION = CMPHA,165,311;%%

\smallskip
\noindent
    Bjerrum-Bohr, N.~E.~J., Dunbar, D.~C., Ita, H., Perkins, W.~B. and Risager, K. (2006):
    {\it The no-triangle hypothesis for $\CN=8$ supergravity},
    JHEP\ {\bf 0612}\ 072
    [arXiv:hep-th/0610043].
    %%CITATION = HEP-TH 0610043;%%

\smallskip
\noindent
    Boels, R., Mason, L.~J. and Skinner, D.\ (2007a)
    {\it Supersymmetric gauge theories in twistor space},
    JHEP\ {\bf 0702}\ 014
    [arXiv:hep-th/0604040].
    %%CITATION = HEP-TH 0604040;%%

\smallskip
\noindent
    \rule{1cm}{.5pt}, \rule{1cm}{.5pt} and \rule{1cm}{.5pt} (2007b):
    {\it From twistor actions to MHV diagrams},
    Phys.\ Lett.\ B\ {\bf 648}\ 90
    [arXiv:hep-th/0702035].
    %%CITATION = HEP-TH 0702035;%%

\smallskip
\noindent
    Cap, A. and Eastwood, M.~G. (2003):
    {\it Some special geometry in dimension six}
    Proc. of the 22nd Winter School {\it Geometry and physics} (Srni 2002),  
    Rend.\ Circ.\ Mat.\ Palermo (2) Suppl.\ No.\ 71\ 93
    [arXiv:math.DG/0003059].

\smallskip
\noindent
    Christensen, S.~M., Deser, S., Duff, M.~J. and Grisaru, M.~T. (1979):
    {\it Chirality, self-duality, and supergravity counterterms},
    Phys.\ Lett.\ B\ {\bf 84}\ 411.
    %%CITATION = PHLTA,B84,411;%%

\smallskip
\noindent
    Dijkgraaf, R., Gukov, S., Neitzke, A. and Vafa, C. (2005):
    {\it Topological M-theory as unification of form theories of gravity},
    Adv.\ Theor.\ Math.\ Phys.\  {\bf 9}\ 603
    [arXiv:hep-th/0411073].
  %%CITATION = 00203,9,603;%%

\smallskip
\noindent
    Green, M.~B., Russo, J.~G. and Vanhove, P. (2007):
    {\it Ultraviolet properties of maximal supergravity},
    Phys.\ Rev.\ Lett.\ {\bf 98}\ 131602
    [arXiv:hep-th/0611273].
    %%CITATION = HEP-TH 0611273;$$

\smallskip
\noindent
    Kallosh, R.~E. (1979):
    {\it Super self-duality},
    JETP\ Lett.\  {\bf 29}\ 172
    [Pisma Zh.\ Eksp.\ Teor.\ Fiz.\ {\bf 29} 192].
    %%CITATION = ZFPRA,29,192;%%

\smallskip
\noindent
    \rule{1cm}{.5pt} (1980):
    {\it Self-duality in superspace},
    Nucl.\ Phys.\  B\ {\bf 165}\ 119.
    %%CITATION = NUPHA,B165,119;%%

\smallskip
\noindent
    Karnas, S. and Ketov, S.~V. (1998):
    {\it An action of $\CN=8$ self-dual supergravity in ultra-hyperbolic harmonic superspace},
    Nucl.\ Phys.\ B\ {\bf 526}\ 597
    [arXiv:hep-th/9712151].
    %%CITATION = HEP-TH 9712151;%%

\smallskip
\noindent
    Ketov, S.~V., Nishino, H. and Gates, S.~J.~J. (1992): 
    {\it Self-dual supersymmetry and supergravity in Atiyah-Ward space-time}, 
    Nucl.\ Phys.\ B\ {\bf 393}\ 149 
    [arXiv:hep-th/9207042].
    See also
    Phys.\ Lett.\ B\ {\bf 297}\ (1992)\ 323 [arXiv:hep-th/9203078],
    Phys.\ Lett.\ B\ {\bf 307}\ (1993)\ 331 [arXiv:hep-th/9203080],
    Phys.\ Lett.\ B\ {\bf 307}\ (1993)\ 323 [arXiv:hep-th/9203081].
    %%CITATION = HEP-TH 9207042;%%
    %%CITATION = HEP-TH 9203081;%%
    %%CITATION = HEP-TH 9203080;%%
    %%CITATION = HEP-TH 9203078;%%

\smallskip
\noindent
    Lechtenfeld, O. and S\"amann, C. (2006):
    {\it Matrix models and $D$-branes in twistor string theory},
    JHEP\ {\bf 0603}\ 002
    [arXiv:hep-th/0511130].
    %%CITATION = JHEPA,0603,002;%%

\smallskip
\noindent
    Manin, Yu.~I. (1988):
    {\it Gauge field theory and complex geometry},
    Springer Verlag, New York 
    [Russian: Nauka, Moscow, 1984].

\smallskip
\noindent
    Mason, L.~J. and Newman, E.~T. (1989):  
    {\it A connection between the Einstein and Yang-Mills equations}, 
    Commun.\ Math.\ Phys.\ {\bf 121}\ 659.

\smallskip
\noindent
%\cite{Mason:2007zv}
%\bibitem{Mason:2007zv}
  Mason, L.\ J.\ and Skinner, D.\ (2008)
  {\it Heterotic twistor-string theory},
  Nucl.\ Phys.\  B {\bf 795} (2008) 105
  [arXiv:0708.2276 [hep-th]].
  %%CITATION = NUPHA,B795,105;%%

\smallskip
\noindent
    \rule{1cm}{.5pt} and Woodhouse, N.~M.~J. (1996):
    {\it Integrability, self-duality, and twistor theory},
    Clarendon Press, Oxford.
    %\href{http://www.slac.stanford.edu/spires/find/hep/www?irn=3540308}{SPIRES entry}

\smallskip
\noindent
    \rule{1cm}{.5pt} (2005):
    {\it Twistor actions for non-self-dual fields: A deriation of twistor string theory},
    JHEP\ {\bf 0510}\ 009
    [arXiv:hep-th/0507269].

\smallskip
\noindent
    \rule{1cm}{.5pt} and Skinner, D. (2006):
    {\it An ambitwistor Yang-Mills Lagrangian},
    Phys.\ Lett.\ B {\bf 636} 60
    [arXiv:hep-th/0510262].

\smallskip
\noindent
    Merkulov, S.~A. (1991):
    {\it Paraconformal supermanifolds and non-standard $\CN$-extended supergravity models},
    Class.\ Quant.\ Grav.\ {\bf 8}\ 557.
    %%CITATION = CQGRD,8,557;%%

\smallskip
\noindent
    \rule{1cm}{.5pt} (1992a):
    {\it Supersymmetric non-linear graviton},
    Funct.\ Anal.\ Appl.\ {\bf 26}\ 69.

\smallskip
\noindent
    \rule{1cm}{.5pt} (1992b):
    {\it Simple supergravity, supersymmetric non-linear gravitons and supertwistor theory},
    Class.\ Quant.\ Grav.\  {\bf 9}\ 2369.
    %%CITATION = CQGRD,9,2369;%%

\smallskip
\noindent
    \rule{1cm}{.5pt} (1992c):
    {\it Quaternionic, quaternionic K\"ahler, and hyper-K\"ahler supermanifolds},
    Lett.\ Math.\ Phys.\ {\bf 25} 7.

\smallskip
\noindent
    Nair, V.P. (2008):
    {\it  A note on graviton amplitudes for new twistor string theories}, 
    Phys.\ Rev.\  D {\bf 78} (2008) 041501
    [arXiv:hep-th/0710.4961].

\smallskip
\noindent
    Penrose, R. (1968):
    {\it Twistor quantization and curved space-time},
    Int.\ J.\ Theor.\ Phys.\ {\bf 1}\ 61.
    %%CITATION = IJTPB,1,61;%%

\smallskip
\noindent
    \rule{1cm}{.5pt} (1976):
    {\it Non-linear gravitons and curved twistor theory},
    Gen.\ Rel.\ Grav.\ {\bf 7}\ 31.
    %%CITATION = GRGVA,7,31;%%

\smallskip
\noindent
    Popov, A.~D. and Wolf, M. (2004):
    {\it Topological B model on weighted projective spaces and self-dual models in four dimensions},
    JHEP\ {\bf 0409}\ 007
    [arXiv:hep-th/0406224].
    %%CITATION = JHEPA,0409,007;%%

\smallskip
\noindent
    \rule{1cm}{.5pt} and S\"amann, C. (2005):
    {\it On supertwistors, the Penrose-Ward transform and $\CN=4$ super Yang-Mills theory},
    Adv.\ Theor.\ Math.\ Phys.\  {\bf 9}\ 931
    [arXiv:hep-th/0405123].
    %%CITATION = HEP-TH 0405123;%%

\smallskip
\noindent
    \rule{1cm}{.5pt}, \rule{1cm}{.5pt} and Wolf, M. (2005):
    {\it The topological B model on a mini-su\-pertwistor space and supersymmetric
    Bogomolny monopole equations},
    JHEP\ {\bf 0510}\ 058
    [arXiv:hep-th/0505161].
    %%CITATION = HEP-TH 0505161;%%

\smallskip
\noindent
    S\"amann, C. (2005):
    {\it The topological B model on fattened complex manifolds and subsectors
    of  $\CN=4$ self-dual Yang-Mills theory},
    JHEP\ {\bf 0501}\ 042
    [arXiv:hep-th/0410292].
    %%CITATION = JHEPA,0501,042;%%

\smallskip
\noindent
    \rule{1cm}{.5pt} (2006):
    {\it Aspects of twistor geometry and supersymmetric field theories within
    superstring theory},
    Ph.D. thesis, Leibniz University of Hannover
    [arXiv:hep-th/0603098].
    %%CITATION = HEP-TH/0603098;%%

\smallskip
\noindent
    Siegel, W. (1992):
    {\it Self-dual $\CN=8$ supergravity as closed $N=2$ ($N=4$) strings},
    Phys.\ Rev.\ D\ {\bf 47}\ 2504
    [arXiv:hep-th/9207043].
    %%CITATION = HEP-TH 9207043;%%

\smallskip
\noindent
    Sokatchev, E.~S. (1995):
    {\it Action for $\CN=4$ supersymmetric self-dual Yang-Mills theory},
    Phys.\ Rev.\ D\ {\bf 53}\ 2062
    [arXiv:hep-th/9509099].
    %%CITATION = PHRVA,D53,2062;%%

\smallskip
\noindent
    Stelle, K.~S. (2006):
    {\it Counterterms, holonomy and supersymmetry}, 
    in {\it Deserfest: A celebration of the Life and works of Stanley Deser, Ann
    Arbor Michigan, 2004}, Eds. J.~T.~Liu, M.~J.~Duff, K.~S.~Stelle,
    R.~P.~Woodward, World Scientific, 303
    [arXiv:hep-th/0503110].
    %%CITATION = HEP-TH 0503110;%%

\smallskip
\noindent
    Waintrob, A.~Yu. (1986):
    {\it Deformations and moduli of supermanifolds}, 
    in {\it Group theoretical methods in physics}, Vol. 1, Nauka, Moscow.

\smallskip
\noindent
    Ward, R.~S. (1980):
    {\it Self-dual space-times with cosmological constants},
    Commun.\ Math.\ Phys.\ {\bf 78}\ 1.

\smallskip
\noindent
    \rule{1cm}{.5pt} and Wells, R.~O. (1990):
    {\it Twistor geometry and field theory},
    Cambridge University Press, Cambridge.
    %\href{http://www.slac.stanford.edu/spires/find/hep/www?irn=2256894}{SPIRES entry}

\smallskip
\noindent
    Witten, E. (1989):
    {\it Topology changing amplitudes in $(2+1)$-dimensional gravity},
    Nucl.\ Phys.\  B\ {\bf 323}\ 113.
    %%CITATION = NUPHA,B323,113;%%

\smallskip
\noindent
    \rule{1cm}{.5pt} (2004):
    {\it Perturbative gauge theory as a string theory in twistor space},
    Commun.\ Math.\ Phys.\  {\bf 252}\ 189
    [arXiv:hep-th/0312171].
    %%CITATION = CMPHA,252,189;%%

\smallskip
\noindent
    Wolf, M. (2006):
    {\it On supertwistor geometry and integrability in super gauge theory},
    Ph.D. thesis, Leibniz University of Hannover
    [arXiv:hep-th/0611013].
    %%CITATION = HEP-TH/0611013;%%

\smallskip
\noindent
    \rule{1cm}{.5pt} (2007):
    {\it Self-dual supergravity and twistor theory},
    Class.\ Quant.\ Grav.\  {\bf 24}\ 6287
    [arXiv:0705.1422 [hep-th]].
    %%CITATION = ARXIV:0705.1422;%%

\smallskip
\noindent
     Woodhouse, N.~M.~J. (1985):
     {\it Real methods in twistor theory},
     Class.\ Quant.\ Grav.\ {\bf 2}\ 257.
     %%CITATION = CQGRD,2,257;%%

\vspace*{1cm}

}

\end{document}